\documentclass[12pt]{article}
\pdfoutput=1
\usepackage[english]{babel}
\newlength{\abstractwidth}
\usepackage{amsfonts}

\usepackage{epsf}
\usepackage{color}
\usepackage{graphicx}
\usepackage{tikz}
\usepackage{amssymb}
\usepackage{latexsym}

\usepackage{color}
\definecolor{dgreen}{rgb}{0,0.70,0.30}
\definecolor{gold}{rgb}{0.85,.66,0}
\definecolor{purple}{rgb}{1.0,0.3,0.6}

\renewcommand{\thefootnote}{\fnsymbol{footnote}}
\renewcommand{\thanks}[1]{\footnote{#1}}
\newcommand{\starttext}{
\setcounter{footnote}{0}
\renewcommand{\thefootnote}{\arabic{footnote}}}

\newcommand{\bea}{\begin{eqnarray}}
\newcommand{\eea}{\end{eqnarray}}
\newcommand{\ba}{\begin{eqnarray}}
\newcommand{\ea}{\end{eqnarray}}
\newcommand{\be}{\begin{eqnarray}}
\newcommand{\ee}{\end{eqnarray}}

\newcommand{\Acknowledgement}[1]{\ \\{\bf Acknowledgements:} #1}

\def\e{\epsilon}

\def\tet{\vartheta}
\def\ep{\varepsilon}

\def\nn{\nonumber}

\def\cA{{\cal A}}
\def\cB{{\cal B}}

\def\cJ{{\cal J}}

\def\cM{{\cal M}}
\def\cN{{\cal N}}
\def\cO{{\cal O}}

\def\cR{{\cal R}}

\def\cT{{\cal T}}

\def\cY{{\cal Y}}

\def\M{{\cal M}}

\def\ba{{\bf a}}

\def\ZZ{{\mathbb Z}}
\def\RR{{\mathbb R}}

\def\Re{{\rm Re \,}}
\def\Im{{\rm Im \,}}

\def\det{{\rm det \,}}

\def\half{ {1\over 2}}
\def\threeh{\frac{3}{2}}
\def\fiveh{\frac{5}{2}}

\def\quart{\frac{1}{4}}
\def\p{\partial}

\def\no{\nonumber}
\def\sm{\smallskip}

\begin{document}
\starttext
\setcounter{footnote}{0}

\begin{flushright}
2013 August 29 \\
DAMTP-2013-43
\end{flushright}

\bigskip

\begin{center}

{\Large \bf Zhang--Kawazumi Invariants  and Superstring Amplitudes }

\vskip .5in

{\large \bf Eric D'Hoker$^{(a)}$ and Michael B. Green$^{(b)}$}

\vskip .2in

{ \sl (a) Department of Physics and Astronomy }\\
{\sl University of California, Los Angeles, CA 90095, USA}

\vskip 0.08in

{ \sl (b) Department of Applied Mathematics and Theoretical Physics }\\
{\sl Wilberforce Road, Cambridge CB3 0WA, UK}

\vskip 0.1in

{\tt \small dhoker@physics.ucla.edu; M.B.Green@damtp.cam.ac.uk}

\end{center}

\vskip .2in

\begin{abstract}

Invariance of Type IIB superstring theory under $SL(2,\ZZ)$ or S-duality implies dependence 
on the complex coupling $T$ through real and complex modular forms in $T$.
Their structure may be understood explicitly in an expansion of superstring corrections 
to Einstein's equations of gravity, in powers of derivatives $D$ and curvature $\cR$.
The perturbative loop expansion in the string coupling 
for the 4-string amplitude governs corrections of the form $D^{2p} \cR^4$ for all $p$.
We show that, at two-loop order, the $D^6 \cR^4$ term is proportional to the integral of 
a {\sl modular invariant} introduced by Zhang and Kawazumi in number theory and related 
to the Faltings $\delta$-invariant studied for genus-two  by Bost. The structure of two-loop 
superstring amplitudes for $p>3$ leads to higher invariants, which generalize  
Zhang--Kawazumi invariants at genus two. An explicit formula is derived for the unique 
higher invariant associated with order $D^8 \cR^4$. In an attempt to compare the prediction 
for the $D^6 \cR^4$ correction from superstring perturbation theory with the one produced 
by S-duality and supersymmetry of Type IIB, various reformulations of the invariant are given.  
This comparison with string theory leads to a predicted value for the integral of the 
Zhang-Kawazumi invariant over the moduli space of genus-two surfaces.

\end{abstract}

\newpage

\tableofcontents

\newpage

\baselineskip=15pt
\setcounter{equation}{0}
\setcounter{footnote}{0}

\section{Overview and outline}
\setcounter{equation}{0}
\label{sec1}

There are a variety of tools for approximating string theory scattering amplitudes.
String perturbation theory is an expansion in powers of the string coupling parameter, $g_s$,
that generalizes the field theoretic Feynman diagram expansion. A term of order $g_s^{2h-2}$ 
in the expansion is referred to an $h$-loop contribution. It arises from integrating over the 
moduli space $\cM_h$ of genus $h$ Riemann surfaces.
Although there is a large body of literature concerning the structure of superstring 
perturbation theory and its effective field theory limits  there are few explicit multi-loop 
amplitude results.  Indeed, the highest order explicit amplitude calculations are at 
two loops, where the four-string amplitude in closed superstring theories has been 
reduced to an integral over the genus-two moduli space $\cM_2$ \cite{D'Hoker:2001nj,D'Hoker:2005jc} 
(see also \cite{D'Hoker:2002gw} for a survey, and references to earlier work, as well 
as \cite{Berkovits:2005ng} for the relation  with the pure spinor approach).  

\sm

An alternative approximation of (super)string amplitudes is the low energy, or $\alpha'$,  
expansion (where $\alpha'$ is the square of the string length scale), in which 
successive terms describe local and nonlocal interactions of higher dimension with the 
lowest order term typically defining a point-like field theory limit based on classical (super)gravity.
Each term in this expansion depends on the moduli, or scalar fields, that characterize the theory.  
Expanding around the boundary of moduli space  gives the perturbation expansions of these coefficients.  
Although the low energy expansion of the tree amplitude is easy to analyse and the one-loop 
amplitude has been studied up to order $(\alpha')^6$, there has been no discussion of the low 
energy expansion of the two-loop amplitude beyond its lowest order non-zero term.

\sm

It is fruitful to consider the constraints imposed by $SL(2,\ZZ)$-duality (which in physics is 
often referred to as $S$-duality) together with supersymmetry on the combination of  the $\alpha'$ 
expansion and string perturbation theory.  Since $SL(2,\ZZ)$-duality relates theories in different 
regions of moduli space it is a non-perturbative feature.   In particular, effective interactions at 
any order in the  low energy  expansion of the amplitude must transform  covariantly under $SL(2,\ZZ)$-duality.  The moduli, or couplings,  dependence of certain  highly supersymmetric  interactions  
that arise at low orders in $\alpha'$ are exactly determined by the $SL(2,\ZZ)$-duality constraints 
and in such cases this leads to precise relationships between perturbative contributions at different orders in perturbation theory. 

\sm

The simplest non-trivial example of $SL(2,\ZZ)$-duality arises in the  ten-dimensional Type IIB theory.
In this theory, the string coupling $g_s$ is related to the imaginary part of a complex 
coupling $T=T_1 + i T_2$ by the relation $T_2 = g_s ^{-1}$ and the requirement $T_2 >0$. The duality group
$SL(2,\ZZ)$ acts on $T$ by M\"obius transformations, and includes exchanges of weak and strong coupling,
namely small and large $g_s=T_2 ^{-1}$. Invariance under $SL(2,\ZZ)$ duality  implies that 
the coefficient of any effective interaction in the low energy expansion of a Type IIB superstring 
amplitude is a function of $T$ that must transform covariantly under $SL(2,\ZZ)$ and encode 
the exact dependence of the interaction on the string coupling.    Interactions of low enough 
dimension satisfy supersymmetry conditions  and a great deal is known about their moduli dependence.  
In particular,   supersymmetry together with $SL(2,\ZZ)$ invariance can be used to determine  
the exact $T$ dependence of the coefficients of the  the first two orders in the low energy 
expansion of the effective action  beyond classical supergravity   
\cite{Green:1997tv,Green:1998by, Sinha:2002zr}. These are the ${\cal R}^4$ interaction 
(which preserves half of the total number of 32 supersymmetries)  and the  $D^4{\cal R}^4$ interaction 
(which preserves 8 supersymmetries).  The quantity $D^{2p} {\cal R}^n$ schematically 
represents a scalar built out of $n$ factors of the Riemann curvature tensor $\cR$ and 
$2p$ covariant derivatives $D$.    In the perturbative limit, $g_s\to 0$, these coefficients 
only contain two perturbative terms, namely a tree-level and a one-loop term (for the $\cR^4$ case) 
or a two-loop term (for the $D^4\,\cR^4$ case).
 
 \sm
 
 The expression for the coefficient  of an interaction preserving only 4 supersymmetries has also 
 been strongly motivated from arguments based on $SL(2,\ZZ)$-duality of M-theory on a torus 
 and is conjectured to satisfy an inhomogeneous Laplace eigenvalue equation in moduli space   
 \cite{Green:2005ba}.  However, this structure has yet to be derived directly by use of supersymmetry.   
 This function possesses four power-behaved terms in its zero Fourier mode, corresponding to 
 string perturbation theory contributions from genus zero to genus three and receives no corrections 
 at higher orders in perturbation theory.  However, only the genus-zero and genus-one components 
 of this coefficient function have been tested by direct comparison with perturbative string amplitude 
 calculations, although there is also indirect evidence that the genus-three component is correct.
 
 \sm
 
Motivated by the preceding comments, in this paper we will initiate the study of the low  energy, or $\alpha'$, 
expansion of the genus-two amplitude by considering the structure of its first non-trivial term, 
which contributes to the $D^6\,\cR^4$ interaction.  This will be expressed as an integral of  an 
$Sp(4,\ZZ)$-invariant over the moduli of the genus-two surface. We will show that this is equal to 
an invariant that has been independently defined in the mathematics literature by Zhang \cite{Zhang} 
and by Kawazumi \cite{Kawazumi}.    This invariant is related \cite{DeJong2} to the Faltings invariant, 
which has special features on genus-two surfaces, as shown by Bost  and collaborators \cite{Bost1,Bost2}.  
Here we will argue that the duality-invariant coefficient of the $D^6\, \cR^4$ interaction  in the 
Type IIB theory gives a prediction for the value of the integral of this invariant over the moduli space of 
genus-two surfaces.  It remains a challenge to perform the integration directly and thereby confirm this  prediction.
 
 \subsection{Outline of paper}
 
The outline of this paper is as follows.   In section~\ref{sec2} we will review the expressions 
for the four-string amplitudes of Type II closed-string theories in superstring perturbation theory up to two loops (up to this order in perturbation theory there is no distinction between Type IIA and Type IIB).  We will describe the structure of the 
low energy expansion of these expressions, which is a sum of powers of Mandelstam invariants.  The expansion of the tree-level (genus-zero) amplitude is straightforward and gives coefficients that are rational numbers multiplying monomials in Riemann zeta values.  The expansion of the genus-one amplitude is more subtle since it involves integrating products of Green functions between points on a given surface, followed by integration over the complex structure. Importantly, the amplitude includes non-analytic  parts that need to be subtracted before expanding the analytic part of the amplitude.    We will survey the structure of the genus-one amplitude before turning to the genus-two case.  

\sm

 The genus-two four-string amplitude is expressed as an integral of the four vertex operator 
 positions on a given Riemann surface $\Sigma$ paramaterized by a period matrix $\Omega$,  followed 
 by integration over $\Omega$ in the moduli space $\cM_2$ of genus 2 Riemann surfaces. 
The leading term in the low energy limit is of order $D^4\, {\cal R}^4$, with a normalization that 
was determined in \cite{D'Hoker:2005ht}, as will also be reviewed in section~\ref{sec2}.   

\sm

The next term in the low energy expansion is of order $D^6\, {\cal R}^4$.  The coefficient of this term, which  is the main focus of interest in this paper, is given by an integral  of a density $\cB _2 ^{(0,1)}(\Omega)$ over genus-two Riemann surfaces parameterised by the period matrix $\Omega$.   In section~\ref{sec3} we will  show that  $\cB _2 ^{(0,1)}(\Omega)$  is given by a certain projection of the scalar Green function, 
\bea
\label{bres}
\cB _2 ^{(0,1)}(\Omega) =  - 8  \int _{\Sigma ^2}   P(z,w) \, G(z,w)\,,
\eea
where the  $P(z,w)$ is a section of $K_z \otimes \bar K_z \otimes K_w \otimes \bar K_w$, and 
$K$ is the canonical bundle on $\Sigma$.  Further manipulations will lead to the identification $\cB _2 ^{(0,1)}(\Omega) = 64\, \varphi(\Omega)$, where  $\varphi$ is  an invariant  that has been considered for altogether different reasons in papers by Zhang \cite{Zhang},  Kawazumi \cite{Kawazumi} and De Jong \cite{DeJong2,DeJong1}.    Generalizations to higher order invariants are obtained in an obvious manner by expanding the string theory $N$-particle amplitude to higher orders in $\alpha'$ as briefly discussed in section~\ref{sec4}.

\sm

In section~\ref{sec5} we will study further properties of $\varphi$, making  use of its relation to the 
Faltings invariant, $\delta$, that was obtained in \cite{DeJong2}.  This leads to an expression 
for $\varphi$ in the form,
\bea
\label{torusint}
\varphi (\Omega) 
=   \varphi_0  - {1 \over 4}  \ln | \Psi _{10} (\Omega)  |^2 + 5 \ln  \Phi  (\Omega)
\eea
where $\varphi_0$ is a simple constant, and $\Psi_{10}$ is the weight-ten Igusa cusp form. 
Also, $\Phi$ is a real-valued  genus-two modular form of weight $(1,1)$ defined by an integral  over the 
real four-dimensional torus $T^4=(\RR/\ZZ)^4$ associated with the Jacobian of the surface,
\bea
\label{phideff}
\ln \Phi (\Omega) =
 \int _{T^4} d^4x  \ln \Big | \tet  [ x  ] ( 0, \Omega ) \Big |^2  \,.
\eea
We will confirm that $\varphi$ is not pluri-harmonic, i.e. it is not the real part of a holomorphic 
function in $\Omega$ (a result shown in \cite{Kawazumi}; see also \cite{DeJong2}),  
by showing that also $\ln \Phi$ is not  pluri-harmonic. The obstruction will be simply related to
the non-trivial dependence, at genus two,  of the $\tet$-divisor on $\Omega$.
An alternative simplified  expression for $\varphi$ is obtained in appendix~\ref{newform}. 

\sm

In section~\ref{sec6} we will discuss the integral of $\varphi(\Omega)$ over moduli space, which is relevant for the connection with the coefficients of the low energy expansion of the string theory amplitude. Although we will prove that this integral is finite (with details given in appendix~\ref{degenerat}) we have not succeeded in evaluating it.  

\sm

We are therefore led in section~\ref{sec7} to consider the value of this integral based on its 
connection to the low energy expansion of Type IIB superstring theory, which is highly constrained 
by $SL(2,\ZZ)$-duality.     We will, in particular,  review the structure of the moduli-dependent 
coefficients of the three leading terms in the $\alpha'$ expansion beyond the classical Einstein  
(super)gravity term, that were mentioned earlier.   The first two of these (the coefficients of $\cR^4$ 
and $D^4\, \cR^4$)  are specific examples of non-holomorphic Eisenstein series, which satisfy 
Laplace eigenvalue equations in moduli space.   The perturbative expansion of such series' 
(i.e., the expansion as $T_2\to \infty$) possess precisely two power-behaved pieces in their zero 
Fourier mode that reproduce the tree-level, genus-one and genus-two parts of these interactions.  
The absence of higher-order corrections to ${\cal R}^4$ beyond genus one and to $D^4 {\cal R}^4$ 
beyond genus two  are striking non-renormalization conditions.        

\sm

 The form of the  coefficient  of the interaction $D^6\,\cR^4$, which preserves 4 supersymmetries, 
 has also been strongly motivated from arguments based on $SL(2,\ZZ)$-duality of M-theory on a torus  \cite{Green:2005ba}, and is conjectured to satisfy an inhomogeneous Laplace eigenvalue equation in moduli space.  The function that satisfies this equation possesses four power behaved terms in its zero Fourier mode, corresponding to string perturbation theory contributions from genus zero to genus three and receives no corrections at higher orders in perturbation theory.  The genus-zero and genus-one contributions have been checked  by direct comparison with perturbative string amplitude calculations.   The genus-three contribution has not been checked directly.  However, an indirect indication that the predicted value of the Type IIB genus-three contribution to $D^6 {\cal R}^4$ is correct is the agreement of its value with the value of the corresponding  Type IIA contribution that was obtained from M-theory compactified on a circle \cite{Green:2005ba}.  
 
 \sm
 
The genus-two contribution to $D^6\,\cR^4$  relates directly to the content of this paper.  
We will show that the value of this contribution contained in the conjectured $SL(2,\ZZ)$-duality invariant 
coefficient leads to a predicted value for the integrated Zhang--Kawazumi invariant,
\bea
\label{result}
\int_{\cM_2} d\mu_2 \, \varphi = \frac{3}{2}\,V_2 = \frac{2 \pi^3}{45} \,,
\eea
where $d\mu_2$ is the $Sp(4,\ZZ)$-invariant measure and $V_2 = \int d\mu_2$ is the volume of 
the moduli space of genus 2 Riemann surfaces $\cM_2$. An explicit check of this relation would 
be of interest, both for its mathematical content and for confirming the $SL(2,\ZZ)$-duality prediction.

\section{Low energy expansion of  Type IIB amplitudes}
\setcounter{equation}{0}
\label{sec2}

The overall kinematic structure of the exact four-string amplitudes are constrained by 
maximal supersymmetry to have the form 
\bea
{\bf A^{(4)}}(\zeta_i,k_i,T)= 
\kappa_{10}^2\, {\bf R}_{\zeta_1,\zeta_2,\zeta_3,\zeta_4}^4(k_{1},k_{2},k_{3},k_{4})\, \cT(s,t,u;T)\,.
\label{kinemamp}
\eea
where 
  \begin{equation}
  {\bf R}_{\zeta_1,\zeta_2,\zeta_3,\zeta_4}^4(k_{1},k_{2},k_{3},k_{4})= \zeta_{1}^{AA'} \zeta_{2}^{BB'}
  \zeta_{3}^{CC'}\zeta_{4}^{DD'}\, K_{ABCD}\,\tilde K_{A'B'C'D'}\,.
  \label{kinfact}
 \end{equation}
 The external states are any of the 256 massless states in the $\cN=2$ supermultiplet of 
 Type IIB  superstring theory, and are described by  polarization tensors,  $\zeta_i^{AB}$ 
 $(i=1,\ldots,4)$, where the indices $A, B$   run over both  vector and spinor values. 
 The tensor $K\, \tilde K$ is defined in \cite{Green:1987sp}.   The amplitudes also depend 
 on the momenta of the external massless states, $k_i^\mu$ ($i=1, \ldots,4$, $\mu=0,1,\ldots,9$), 
 which  satisfy $k_i \cdot k_i=0$, and overall momentum conservation requires $k_1+k_2+k_3+k_4=0$.  
It will be convenient to introduce dimensionless Lorentz-invariant variables $s,t,u$ defined by  
$s=-\alpha' (k_1+k_2)^2/4$, $t=-\alpha' (k_2+k_3)^2/4$, $u= -\alpha' (k_1+k_3)^2/4$, and which
obey $s+t+u=0$.  The scalar function $\cT(s,t,u;T)$ in (\ref{kinemamp}) depends on 
$s,t,u$ and the modulus field, $T$.   
 
\subsection{Structure of the full amplitudes} 
 
For convenience, we will follow the notation of \cite{D'Hoker:2001nj,D'Hoker:2005jc} in the construction 
of the amplitudes, which concentrated on the sector of amplitudes with external  NS-NS bosons, 
with polarization tensors $\e _i ^{\mu \bar \mu}$.   Such amplitudes  will be denoted by $\cA^{(4)} (\e_i, k_i, T)$.  
 Since these amplitudes are linear in each $\e^{\mu \bar \mu}_i$, a general amplitude is a linear 
 combination of a basis of amplitudes in which
the polarization tensor is factorized, $\e_i ^{\mu \bar \mu} = \e _i ^\mu \, \bar \e ^{\bar \mu} _i$.  
 More explicitly, the prefactor that multiplies the amplitude has the form 
\bea
\label{2a1}
K \bar K = 2^6 \, \cR^4 
\eea
The  kinematic factor $K$ is normalized as follows,
\bea
\label{2a2}
K 
& = &
(f_1 ^{\mu \nu} f_2^{\nu \mu} ) (f_3^{\rho \sigma}  f_4^{\sigma \rho}) + 
(f_1 ^{\mu \nu} f_3^{\nu \mu} ) (f_2^{\rho \sigma}  f_4^{\sigma \rho}) + 
(f_1 ^{\mu \nu} f_4^{\nu \mu} ) (f_2^{\rho \sigma}  f_3^{\sigma \rho}) 
\no \\ &&
- 4 f_1 ^{\mu \nu} f_2 ^{\nu \rho} f_3^{\rho \sigma}  f_4^{\sigma \mu}  
- 4 f_1 ^{\mu \nu} f_3 ^{\nu \rho} f_2^{\rho \sigma}  f_4^{\sigma \mu}  
- 4 f_1 ^{\mu \nu} f_2 ^{\nu \rho} f_4^{\rho \sigma}  f_3^{\sigma \mu}  
\eea
where we use the following notation for the gauge invariant field strength, 
$f_i ^{\mu \nu} = \e_i ^\mu k_i ^\nu - \e_i ^\nu k_i ^\mu$.
The kinematic factor $\bar K$ is obtained from $K$ by substituting $\e^\mu _i \to \bar \e _i ^{\bar \mu}$.
In the case of four external gravitons the prefactor  ${\cR}^4$ reduces to  the product of four linearized Weyl curvatures  contracted into each other by a well-known sixteen-index  tensor, $t_{8}t_{8}$. 

\sm

In string perturbation theory the amplitude has an expansion in integer powers of 
$T_2^{-1}= g_s$ that has the form
\bea
\label{pertexpand}
\cA^{(4)}(\epsilon_i,k_i,T )\Bigl|_{pert.} =\sum_{h=0}^\infty \cA_h ^{(4)} (\e_i,k_i, T_2) \,,
\eea 
where $\cA_h ^{(4)} (\e_i,k_i,T_2)$ is the $h$-loop amplitude defined by a 
functional integral over genus-$h$ Riemann surfaces, and is proportional 
to $T_2^{2-2h}=g_s^{2h-2}$.  Note that the perturbative  terms in the IIB 
theory do not involve the Ramond--Ramond scalar, $T_1$, but it enters into 
the non-perturbative contributions to the amplitude through the effects of 
D-instantons, as will be apparent when we consider the implementation of 
$SL(2,\ZZ)$ duality later in this paper.   
The properly normalized perturbative amplitudes for $h =0,1,2$ are given as 
follows \cite{D'Hoker:2005ht},\footnote{In the two-loop amplitude $\cA ^{(4)}_2$ given in 
formula (2.23) of  \cite{D'Hoker:2005ht}, it is understood that each factor of $\cY_S$
is accompanied by a factor of $\alpha'/2$, since the convention $\alpha '=2$ was
used in \cite{D'Hoker:2005jc} where these formulas were originally obtained.
Properly restoring these factors produces a factor of 4, which has been carefully
taken into account in writing formula (\ref{gen2amp}) below. We take this opportunity to 
also correct a typo on the last line of equation (2.31) of \cite{D'Hoker:2005jc}, where the 
factor of $\rho$ should be removed.}
\bea
\label{2a4}
\cA_0 ^{(4)} (\e_i,k_i, T_2)
& = & \kappa _{10}^2 \, T_2^2 \, \cR^4
{ \Gamma (- s )\Gamma (-  t )\Gamma (-  u)
\over
\Gamma (1+ s)\Gamma (1+t)\Gamma (1+ u)}\,,
\eea
\bea
\label{22a4}
\cA_1 ^{(4)} (\e_i, k_i,T_2)
& = &
{ \pi \over 16} \kappa _{10}^2 \,  T_2 ^0 \, \cR^4 
\int _{\M_1} {|d\tau|^2 \over ( \Im \, \tau)^2} \, 
\cB_1 (s,t,u| \tau)\,,
\eea
\bea
\label{gen2amp}
\cA_2 ^{(4)} (\epsilon_i, k_i,T_2)
& = &
{ \pi \over 64} \kappa _{10}^2 \, T_2^{-2} \, \cR^4 
\int_{\M_2} { |d^3 \Omega|^2 \over (\det \Im \Omega)^3} \,
\cB_2 (s,t,u | \Omega)
\eea
In these formulas, $\kappa _{10}^2$ is the 10-dimensional Newton constant.   
The dimensionless reduced amplitudes $\cB_h $ at fixed moduli  
only depend on the Mandelstam variables, and are given by,
\bea
\label{2a5}
\cB_1 (s,t,u | \tau)
 = 
\int _{\Sigma^4} {\prod _{i=1}^4 d^2z_i \over (\Im \, \tau)^4}
\exp \left \{ -  \frac{\alpha'}{2}\sum _{i<j} k_i \cdot k_j  \, G(z_i,z_j) \right \}
\\
\label{2a55}
\cB_2  (s,t,u | \Omega)
 =   \int_{\Sigma^4}
{ |{\cal Y}_S|^2 \over (\det \Im \Omega )^2}
\exp \left  \{ -\frac{\alpha'}{2} \sum_{i<j}k_i\cdot k_j\,G(z_i,z_j) \right \} 
\eea
The integration over $\Sigma ^4$ stands for a 4-fold integral over the Riemann surface $\Sigma$. 
To define the other ingredients, we fix a canonical homology basis of 1-cycles $A_I, B_I$ 
with $I=1,\cdots, h$ (with $h=1,2$ in this paper), and a dual basis of holomorphic 1-forms $\omega _I$ satisfying, 
\bea
\oint _{A_I} \omega _J = \delta _{IJ} 
\hskip 1in 
\oint _{B_I} \omega _J = \Omega _{IJ}
\eea
For $h=1$, the holomorphic Abelian differential is constant, $\omega_1 (z) = dz$ in terms 
of a local complex coordinate $z$. The moduli space $\cM_1$ of genus-one Riemann 
surfaces  is parametrized by the local complex  coordinate $\tau=\Omega _{11}$ in the 
range $1 \leq |\tau|$  and $-1 \leq 2\Re (\tau) \leq 1$.  

\sm

For $h=2$, the moduli space $\cM_2$ of genus-two Riemann surfaces is parametrized by the 
entries of the period matrix $\Omega_{IJ}$, subject to the following  set of  inequalities \cite{Klingen},
\bea
\label{domain}
(1) && 0 \leq | 2 \Im (\Omega _{12}) | \leq \Im (\Omega _{11}) \leq \Im (\Omega _{22})
\no \\
(2) & \hskip 1in &
| \Re (\Omega_{11}) | \leq \half, ~ | \Re (\Omega_{22}) | \leq \half, ~ | \Re (\Omega_{12}) | \leq \half
\no \\ 
(3) && |\det ( C \Omega +D) | \geq 1 ~ \hbox{for all} ~ \left ( \matrix{A & B \cr C & D \cr } \right ) \in Sp(4,\ZZ)
\eea
The dependence on moduli of Abelian differentials, the prime form, and the 
Green function will  not exhibited, unless otherwise indicated.
The differential form $\cY_S$ on $\Sigma ^4$ is given by,
\bea
\label{2a6}
3 \cY_S & = &
~~ (t-u) \Delta (1,2) \wedge \Delta (3,4)
\no \\ && 
+ (s-t) \Delta (1,3) \wedge \Delta (4,2)
\no \\ &&
+ (u-s) \Delta (1,4) \wedge \Delta (2,3)
\quad
\eea
where the bi-holomorphic  form $\Delta(z,w)$ is a section of $K_z\otimes K_w$, and is defined by
\bea
\label{2a7}
\Delta (i,j) = \Delta (z_i, z_j) = 
\omega _1(z_i) \wedge \omega _2 (z_j) - \omega _2 (z_i) \wedge \omega _1(z_j) 
\eea
The differential is symmetric $\Delta (j,i) = \Delta (i,j)$, and satisfies the relation,
\bea
\label{2a71}
\Delta (1,2) \wedge \Delta (3,4) + \Delta (1,3) \wedge \Delta (4,2) + \Delta (1,4) \wedge \Delta (2,3) =0
\eea
With the help of (\ref{2a71}), and momentum conservation,
 the following alternative expressions for  $\cY_S$ may be derived,
\bea
\label{2a8}
\cY_S & = & -s \Delta(1,4) \wedge \Delta(2,3) + t \Delta(1,2) \wedge \Delta(3,4)
\no \\
\cY_S & = & -u \Delta(1,2) \wedge \Delta(3,4) + s \Delta(1,3) \wedge \Delta(4,2)
\no \\
\cY_S & = & -t \Delta(1,3) \wedge \Delta(4,2) + u \Delta(1,4) \wedge \Delta(2,3)
\eea
Finally,  for genus one and two,  $G(z,w)$ is a scalar Green function. 
Since  the range of the scalar Laplace operator on a compact Riemann surface is 
orthogonal to the constant function, the scalar Green function is not uniquely defined. 
This non-uniqueness  is reflected in the fact that
one may shift $G$ by an arbitrary function $f$ as follows $ G(z,w) \to G(z,w) + f(z) + f(w)$. 
This shift is inconsequential in the string amplitudes of (\ref{2a4})
in view of momentum conservation, $s+t+u=0$. One convenient choice for the 
Green function is given by, 
\bea
\label{2a9}
G(z,w)=-\ln |E(z,w)|^2+2\pi ({\rm Im}\,\Omega)_{IJ}^{-1} \,
\biggl ( {\rm Im}\int_z^w\omega_I \biggr )
\biggl ( {\rm Im}\int_z^w\omega_J \biggr )
\eea
where $E(z,w)$ is the prime form. For $h=1$, the prime form is given in terms of the Jacobi 
$\tet$-function $\tet_1(z,\tau) = \tet [\half \half ](z,\tau)$ by $E(z,w) = \tet _1 (z-w)/\tet ' _1(0)$ 
for modulus $\tau$, where Jacobi $\tet$-functions with general real characteristics 
$\kappa=[\kappa ' \kappa '']$  are defined by,
\bea
\label{2a11}
\tet [\kappa ' \kappa '']  (z,\tau) = \sum _{n \in \ZZ} \exp \left \{ i \pi \tau (n+k')^2 + 2 \pi i (n+k') (z+ \kappa '') \right \}
\eea
and the Green function takes on a simplified form,
\bea
\label{2a10}
G(z,w) = - \ln \left | { \tet _1 (z-w) \over \tet ' _1(0)} \right |^2 + {2 \pi \over \Im \tau} \left ( \Im (z-w) \right )^2
\eea
For $h=2$, the prime form may be found in (\ref{5a19}) of this paper,
and in equation (3.9) of \cite{D'Hoker:2005jc}, but its 
explicit expression will not be needed here.

\subsection{Structure of the low energy expansion}

For fixed moduli $\tau$ and $\Omega$, the integrations over $\Sigma$ in
the reduced amplitudes $\cB_1$ and $\cB_2$ of  (\ref{2a5}) and (\ref{2a55}) will not converge for all
values of $s,t,u$. Instead, poles will be produced at positive integer values of  $s , t$ 
and $u$. The physical origin of these poles is the appearance of massive on-shell intermediate  
states, just as was the case in the tree-level amplitudes.  Since no such poles, or any other singularities,
can occur for sufficiently small $s,t,u$, the Taylor series expansion in these variables has finite coefficients,
and the series will have a finite radius of convergence.   A separate issue, which arises 
upon further integration over moduli,  is the fact that the loop 
amplitude has non-analytic thresholds, as prescribed by unitarity.  These arise from 
degenerations of Riemann surfaces at  boundaries of moduli space, and were discussed in the context of the 
genus-one case in \cite{Green:1999pv, Green:2008uj}.  Earlier discussions of the analytic
behavior of the one loop amplitude may be found in  \cite{Green:1981yb,D'Hoker:1994yr}.

\sm

Exploiting the invariance of the integrands in $\cB_1$ and $\cB_2$ under permutations of  
the index $i$ on the variables $(z_i, k_i)$,
the Taylor series expansions of the functions $\cB_1$ and $\cB_2$ may be arranged in  symmetric 
polynomials in $s,t,u$.  To do so, we write the exponential factor in the integrals in terms of $s,t,u$, 
\bea
\label{4b1}
\exp \Big \{  sG(1,2) +  tG(1,4)+ uG(1,3) + 
 sG(3,4) +  tG(2,3)+ uG(2,4) \Big \} 
\eea
Since we have $s+t+u=0$, only two independent invariants remain, 
\bea
\label{4a2}
\sigma _2 =  (s^2 + t^2 +u^2) 
\hskip 0.7in 
\sigma _3 = (s^3 + t^3 +u^3) = 3 stu
\eea
Thus, $\cB_h$  will admit the following expansions, 
\bea
\label{4a3}
\cB_h (s,t,u | \Omega) = \sum _{p,q=0}^\infty \cB_h ^{(p,q)} (\Omega) 
\times { (\sigma _2)^p (\sigma _3)^q \over p! \, q!}
\eea
where we will set $\Omega = \tau$ for $h=1$. By construction, the coefficients $\cB_h ^{(p,q)} (\Omega)$
are smooth real modular invariants, and thus depend only on the surface $\Sigma$, and not on the 
specific period matrix representing $\Sigma$.

\sm

However, care has to be taken in integrating the coefficients $\cB _h ^{(p,q)}(\Omega)$  
over moduli space since such integrals may be divergent, just as had already been the case 
for the full superstring amplitudes. The divergent parts are accounted for 
by the presence of non-analytic contributions in the variables $s,t,u$ due to thresholds that are 
prescribed by unitarity \cite{Green:1999pv, Green:2008uj}.  

\subsection{Review of  genus-zero and genus-one  expansions}
\label{zeroone}

Since we will be interested in comparing the coefficients of the $\sigma_3\, R^4$ interaction 
at different genera,   we will here review the low energy expansions up to this order at genus 
$0$ and $1$ before considering the genus $2$ case.

\subsubsection{The genus-zero expansion}

The genus-zero four point amplitude, (\ref{2a4}), can easily be expanded to all orders in the limit of $s, t, u \ll 1$ using standard properties of the $\Gamma$ function.   The first few terms in the expansion are as follows, 
\bea
\label{treeexpand}
\cA_0 ^{(4)} (\e_i,k_i,T_2)
& = & \kappa _{10}^2 \, T_2^2 \, \cR^4 \frac{1}{stu} 
\exp\left(\sum_{n=1}^\infty \frac{2\zeta(2n+1)}{2n+1}(s^n+t^n+u^n)  \right) 
\nn\\
&=&  \kappa _{10}^2 \, T_2^2 \, \cR^4 \left(2\zeta(3) + \zeta(5)\, \sigma_2 
+ \frac{2}{3}\zeta(3)^2 \, \sigma_3 + \dots \right)\,.
\eea
In writing this we have used the fact that \cite{Green:1999pv}
\bea
s^n+ t^n+ u^n =  n \sum_{2p+3q=n}\frac{(p+q-1)!} {p! \, q!}\,
\left(\frac{\sigma_2}{2}\right)^p\, \left(\frac{\sigma_3}{3}\right)^q\,.
\label{mdefs}
\eea
The coefficient of the term of order $\sigma_2^p\, \sigma_3^q\,\cR^4 \sim s^{2p+3q}\,\cR^4$ in this expansion is a monomial in Riemann $\zeta$ values of depth $2p+3q+3$ with rational coefficients\footnote{In the generalisation to the expansion of $N$-particle closed superstring tree amplitudes the coefficients are generally multi-zeta values \cite{Schlotterer:2012ny}.}.

\subsubsection{The genus-one expansion}
\label{genone}

The low energy expansion of loop amplitudes is considerably more difficult than the tree-level case.  
At genus one and higher qualitatively new issues arise since the $Sp(2h,\ZZ)$-invariant coefficients  
$\cB_h^{(p,q)}  (s,t,u| \Omega)$ in (\ref{4a3}) are integrals of  $(2p+3q)$ powers of the Green function on a genus-$h$ surface that arise in the expansion of the exponential factor (\ref{4b1}).  We will here review the genus-one expansion, which was discussed in detail in \cite{Green:1999pv,Green:2008uj}, where  the $SL(2,\ZZ)$-invariant  expansion coefficients were determined up to order $s^6\, \cR^4$.   The genus-one Green function $G(z,w)$
of (\ref{2a10}) may be expressed as a double Fourier expansion in the form,
\bea
\label{newprop}
G(z,w)
=  {1\over \pi}  \sum_{(m,n)\neq(0,0)}
{\tau_2\over |m\tau+n|^2} \exp\left[2\pi  i (nx  - n y)\right] + 2\, \ln\left(2\pi \left|\eta(\tau) \right|^2\right)\,.
  \eea
The Dedekind eta-function $\eta (\tau)$ is defined by,
 \bea
 \eta (\tau) = e^{i \pi \tau/12} \prod _{n=1}^\infty \left ( 1 - e^{2 \pi i n \tau} \right )\, ,
 \eea
 and  we have parametrized  $z-w$ by real coordinates, $x$ and $y$,  
 \bea
 z-w  =x +\tau y\, ,
 \label{doubleperiod}
 \eea 
 so that $x$ and $y$ are normalized to have period $1$.  The zero mode in (\ref{newprop}) 
 (the last term) cancels in the combination of Green functions that arises in the amplitude in 
 the expansion of (\ref{4b1}).    This has the immediate consequence that the  term linear in 
 $G$ does not arise in the expansion (\ref{4a3}).  In this way we may identify the momentum 
 space Green function as,
 \bea
 \label{momengreen}
 \hat G(m,n) =    \frac{1}{\pi} \sum_{(m,n)\neq(0,0)}
{\tau_2\over |m\tau+n|^2} \,,
 \eea
which only contains non-zero modes.

\sm

The coefficient, $\cB_1^{(p,q)}$, of the order $s^{2p+3q}\, R^4$ contribution to the expansion involves sums of terms that are products of $2p+3q$ Green functions joining pairs of vertex positions, which are then integrated over the torus.   Any such term  can be simply expressed in momentum space by a diagram with the four external vertices represented by nodes and each Green function by a propagator joining two of the nodes. The integer world-sheet momenta in each propagator of the form (\ref{momengreen}) are summed with momentum conserved at each vertex.   The absence of a zero momentum component in the propagator (\ref{momengreen}) means that there are no diagrams in which any vertex  has a single propagator joined to it.  In particular, this means that there is no contribution with a single power of the Green function.   This contrasts with the situation at higher genus, where there is a  contribution with a single Green function, which we will consider in detail later.

\sm

The first term in the expansion is the trivial term  with coefficient $B_1^{(0,0)}=16$ in (\ref{4a3})\footnote{Note that with our conventions $\int |dz|^2 = \int |dz\wedge d\bar z| = 2\Im \tau$.}.  Substituting in (\ref{22a4}) gives the leading contribution to the genus-one amplitude, which is proportional to the volume of ${\cal M}_1$, 
\bea
\label{oneleading}
{\cal A}_1^{(4)}(\epsilon_i,k_i,T_2) \Big |_{s,t,u=0}= \frac{2\pi^2}{3}\kappa_{10}^2 {\cal R}^4 \,.
\eea
The first non-trivial term in the expansion is of order $s^4 R^4$ with a coefficient that is proportional to\footnote{Note that the normalization of $\hat G$  in (\ref{momengreen}) differs by a factor of $4\pi$ from that in  \cite{Green:2008uj} and  the definition of $E_s$ in (\ref{dtwodef}) differs by a factor of $2\zeta(2s)$ from the definition in \cite{Green:2008uj}. This leads to differences in the normalizations of the modular invariant coefficients.}
\begin{center}
\centering
\vskip -0.8cm
\begin{tikzpicture} [scale=1.5,line width=0.30mm]\scope[xshift=2.5cm]
\begin{scope}[xshift=1cm,yshift=1cm]
\draw (2,0.5) node{$\bullet$} ..controls (2.5,0.2) ..  (3,0.5) node{$\bullet$};
\draw (2,0.5) node{$\bullet$} ..controls (2.5,0.8) ..  (3,0.5) node{$\bullet$};
\draw(4.2,1.5)[yshift=-1cm] node{$=:\ \frac{2}{\pi^2} \,\zeta(4)\,  \ E_2 \ ,$};
\end{scope}
\endscope\end{tikzpicture}
\label{secondone}
\end{center}
where $E_2(\tau)$ is the $s=2$ case of a  non-holomorphic Eisenstein series, defined by
\bea
E_s(\tau)  =\frac{1}{2\zeta(2s)\, } \sum_{(m,n)\ne (0,0)} \frac{\tau_2^s}{|m + n\tau|^{2s}}=  \sum_{p,q\in \ZZ \atop{gcd(p,q)=0}} \frac{\tau_2^s}{|p + q\tau|^{2s}}\,,
\label{dtwodef}
\eea
which is easily seen to be invariant under $SL(2,\ZZ)$ transformations that act on $\tau$ by
\bea
\label{upsact}
\tau  \to \frac{a\tau  + b}{c\tau + d}\,,\qquad\qquad a,b,c,d \in \ZZ\,  \qquad ad-bc=1\,.
\eea
It also satisfies the Laplace eigenvalue equation 
\bea
\label{laplaceeigenvalue}
\Delta_\tau E_s(\tau) = s(s-1)\, E_s(\tau)\,,
\eea
where the $SL(2)$ Laplace operator is defined by $\Delta_\tau = \tau_2^2(\partial_{\tau_1}^2  +\partial_{\tau_2}^2)$.
The integral of an Eisenstein series over a fundamental $SL(2,\ZZ)$  domain is generally divergent at the boundary $\tau_2\to \infty$ so we will integrate over the cutoff  fundamental domain ${\cal F}_L$ defined by 
\bea
\label{cutofff}
{\cal F}_L = \{\tau | -1/2 \le \tau_1\le 1/2,\   \tau_2 \le L,\ |\tau| \ge 1, \ L \gg 1\}\,.
\eea
Such an  integral is evaluated by using Gauss's law to localize the result on the boundary 
of the cutoff fundamental domain,
\bea
\label{intofeisen}
\int_{{\cal F}_L} \frac {d^2\tau}{\tau_2^2} \, E_s(\tau) =\frac{1}{s(s-1)}\, \int_{{\cal F}_L} \frac {d^2\tau}{\tau_2^2} \, \Delta_\tau E_s(\tau) =
\frac{2\zeta(2s)}{s-1}\, L^{s-1} +O(L^{-s}) \,
\eea 
where we have used the asymptotic behavior of the Eisenstein series, 
$\lim_{\tau_2\to \infty} E_s(\tau) = 2\zeta(2s) \tau_2^s + \cO(\tau_2^{1-s})$.    
Terms that are power behaved in the cutoff $L$ are cancelled once the 
non-analytic part of the amplitude is taken into account. 
The non-analytic contributions arise from the large-$\tau_2$ boundary of (\ref{2a5}).  
In order to isolate these contributions it is necessary to consider the region 
of the integral with $L\le \tau_2 \le \infty$.   The first of these arises at order 
$\cO(s \cR^4\, \ln s)$ and is identified with the logarithmic singularity that can be 
obtained by dimensional regularization of one-loop supergravity. 

\sm

Since the expression for $\int_{{\cal M}_1} E_2(\tau)$ vanishes after subtracting the term linear in $L$ in (\ref{intofeisen}) there is no genus-one contribution to the terms of order $\sigma_2\,  {\cal R}^4$  \cite{Green:1999pv}.  This fits in with expectations based on $SL(2,\ZZ)$-duality that predict that the $\sigma_2 \, {\cal R}^4$ term is absent at genus one but is present at genus two, as will be reviewed in section~\ref{sec7}.

\sm

The two diagrams that contribute to $\cB_1^{(0,1)}$, the coefficient of the term of order $\sigma_3{\cal R}^4$, are
\vskip 0.8cm
\begin{center}
\centering
\vskip -0.8cm
\begin{tikzpicture} [scale=1.5,line width=0.30mm]\scope[xshift=2.5cm]
\begin{scope}[xshift=1cm,yshift=1cm]
\draw (2,-0.5) node{$\bullet$} ..controls (2.5,-0.8) ..  (3,-0.5) node{$\bullet$};
\draw (2,-0.5) node{$\bullet$} ..controls (2.5,-0.2) ..  (3,-0.5) node{$\bullet$};
\draw (2,-0.5) -- (3,-0.5)  ;
\draw(4.0,0.5)[yshift=-1cm] node{$=: \ D_3(\tau) \,.$};
\end{scope}
\endscope
\begin{scope}[xshift=2cm]
\draw (0,0) node{$\bullet$} ;
\draw (0,1) node{$\bullet$} ;
\draw (0.7,0.5) node{$\bullet$} ;
\draw (0,0) -- (0,1) ;
\draw (0,0) -- (0.7,0.5);
\draw (0,1) -- (0.7,0.5);
\draw(1.9,0.5) node{$= \ \frac{2}{\pi^3}\zeta(6)\, E_3(\tau)\,,$};
\end{scope}
\end{tikzpicture}
\label{thirdone}
\end{center}
The first term is another Eisenstein series that gives zero contribution to $\int_{\cM_1} \cB_1^{(0,1)}$  
by the same reasoning as in the earlier case.  However, the coefficient $D_3$ is tricker to evaluate 
and has the form \cite{Green:2008uj}\footnote{This expression was originally believed 
\cite{Green:2008uj} to be an approximation up to terms that vanish in the limit $\tau_2\to \infty$, 
but was subsequently shown to be exact by Zagier (private communication).}
\bea
\label{dthree}
D_3(\tau) = \frac{2}{\pi^3}\zeta(6)\, E_3(\tau)  +\zeta(3)\,.
\eea
Taking into account the combinatorial factor that specifies the number of ways in which the diagram $D_3$ arises from expansion of the exponential (\ref{4b1})  and
 performing the $\tau$ integral over the cutoff fundamental domain (again dropping terms that are power behaved in the cutoff $L$) gives a contribution to the amplitude at order $\sigma_3\, {\cal R}^4$ \cite{Green:1999pv}
\bea
\label{d6one}
{\cal A}_1^{(4)}(\epsilon_i,k_i, T_2) \Big |_{\sigma_3}= \frac{2\pi^2}{9} 
\zeta(3)\,\kappa_{10}^2\, \sigma_3\, {\cal R}^4 \,.
\eea
Note that higher order diagrams contributing to the expansion of the loop amplitude integrand for the $N$-particle amplitude give invariants of the form,
\bea
D_{l_{12},l_{13}, \ldots} (\tau)= \sum_{l_{ij}} \prod_{1\le i<j\le N}  
\frac{\tau_2^{l_{ij}}}{|m_{ij} + n_{ij}\tau|^{2l_{ij}}} \, 
\prod_{i=1}^N \delta \Big ( \sum_j \sigma _{ji}  m_{ij} \Big ) \delta \Big ( \sum_j  \sigma _{ji} \, n_{ij} \Big ) \,.
\label{genformd}
\eea
where $\sigma _{ji}= {\rm sign} (j-i)$, while $l_{ij}$ is the number of propagators joining 
vertices labelled $i$ and $j$, and the weight, $w=\sum_{1\le i<j\le N} l_{ij}$ labels the 
order in the $\alpha'$ expansion.   The Kronecker delta's impose conservation of the 
integer momenta at each vertex labelled by $i$.  In the case of the four-string 
amplitude ($N=4$) diagrams of the form  (\ref{genformd}) arise at order $s^w \cR^4$. 
Some of these higher-order terms were analyzed in   \cite{Green:2008uj}, but we will 
not consider them further here  since they are not of direct relevance to this paper.

\sm

Generalizing to $N$-particle amplitudes with $N>4$ not only leads to analogous diagrams 
with $N$ vertices, but also to modifications of the rules in (\ref{genformd})  to account for 
world-sheet  propagators with numerator momentum factors \cite{Green:2013bza}.

\subsection{The two lowest-order genus-two contributions}

Since the prefactor,  $|\cY_S|^2$, in the  genus $h=2$ amplitude is of degree 2 in $s,t,u$, 
it follows immediately that  $\cB_2 ^{(0,0)}(\Omega)=0$, a result first proven in \cite{D'Hoker:2005jc}.

\sm

The simplest non-zero contribution arising at two-loop level is $\cB_2 ^{(1,0)}$.
It is obtained by retaining the lowest order contribution of  the exponential, namely 1,
and  setting $t=-s$ and $u=0$. Using the Riemann bilinear 
relation for the period matrix $\Omega$, 
\bea
\label{2c1}
{ i \over 2} \int _\Sigma  \omega_I \wedge \overline{\omega_J} =  \Im \Omega _{IJ}
\eea
we readily derive the following expression, 
\bea
\label{2c2}
\cB_2 ^{(1,0)} (\Omega) = \half \int_{\Sigma^4}
{ | \Delta (1,3) \wedge \Delta (2,4) |^2 \over (\det \Im \Omega )^2}=32
\eea
Its value was used in \cite{D'Hoker:2005ht} to compute  the coefficient of the correction $D^4\cR^4$
to two loop order, giving the result,
\bea
\label{d6two}
\cA_2 ^{(4)} (\epsilon_i, k_i,T_2)\Big |_{\sigma_2}
=
\frac{\pi}{2} V_2 \kappa _{10}^2 \, T_2 ^{-2} \, \sigma_2 \cR^4 
=
\frac{2\pi ^4}{135}  \kappa _{10}^2 \, T_2 ^{-2} \, \sigma_2 \cR^4 \,
\eea
We have used the fact that the volume of $\cM_2$ is $V_2=4\pi^3/135$
(see for example \cite{Klingen} and Appendix A of \cite{D'Hoker:2005jc}).
As we will review later, this value is in precise agreement with the one expected from the 
implementation of $SL(2,\ZZ)$-duality at order $\sigma_2\, {\cal R}^4$.

\section{Relating $\cB _2 ^{(0,1)}$ to the Zhang--Kawazumi invariant}
\setcounter{equation}{0}
\label{sec3}

We will now simplify the first non-trivial term in the expansion of the genus-two amplitude, 
which has the form  $ D^6 \cR^4 \int_{\cM_2} d\mu_2\,\cB_2 ^{(0,1)}$.   
This is the term that is linear in the Green function~$G$. We shall then review the 
definition of an invariant introduced by Zhang \cite{Zhang} and by Kawazumi  
\cite{Kawazumi}, and show that for genus two  it is proportional to $\cB_2 ^{(0,1)} (\Omega)$.

\subsection{Simplification of $\cB_2 ^{(0,1)}$}

Given the general expansion of $\cB_2(s,t,u | \Omega )$ in terms of $s,t,u$,
we may set $s=t$ and $u=-2s$ to determine $\cB_2 ^{(0,1)} (\Omega)$, 
while choosing the $s,t$ symmetric representation for $\cY_S$ on the first line of (\ref{2a8}). We find
the following expression,
\bea
\label{3a1}
\cB_2 ^{(0,1)} (\Omega) & = & - { 1 \over 3} \int _{\Sigma ^4}
{ |\Delta (1,2) \wedge \Delta (3,4) - \Delta (1,4) \wedge \Delta (2,3)|^2 \over (\det Y )^2}
\no \\ && \hskip 0.4in \times
\Big \{  G(1,2)+ G(3,4)  -G(1,3) - G(2,4) \Big \} 
\eea
where we shall use the abbreviation $Y = \Im \Omega$ throughout.
The term involving $G(1,2)$ may be integrated over the points 3 and 4, and so on,  
making use of  the following formulas, 
\bea
\label{3a2}
\int _{\Sigma _i} \Delta (i,j) \wedge \overline{\Delta (i,k)}  & = & 
2 i \, (\det Y) \sum _{J,K} Y^{-1} _{JK}  \omega_J(j) \wedge \overline \omega_K(k)
\no \\
\int _{\Sigma _j} \int _{\Sigma _k} \Delta (i,j) \wedge 
\overline{\Delta (j,k)} \wedge \Delta (k, \ell) & = &- 4 \, (\det Y) \, \Delta (i,\ell)\,
\eea
which follow from  (\ref{2c1}).  As a result, we find, 
\bea
\label{3a3}
\cB _2 ^{(0,1)}(\Omega) =  - 8  \int _{\Sigma ^2}   P(z,w) \, G(z,w)
\eea
We have introduced the form $P(z,w)$ of tensor type $(1,1)_z \otimes (1,1)_w$,
which may be defined for arbitrary genus $h$ by, 
\bea
\label{3a4}
P(z,w) = \sum _{I,J,K,L} \left ( -Y^{-1}_{IJ} Y^{-1}_{KL}  + h Y^{-1}_{IL}  Y^{-1}_{JK}  \right ) 
  \omega_I(z) \wedge \overline{\omega_J(z)}  \wedge \omega_K(w) \wedge \overline{\omega_L(w)} 
\eea
It is readily verified that $P(z,w)$ is symmetric under interchange of $z$ and $w$,
and  integrates to 0 against a constant function,
\bea
\label{3a5}
\int _{\Sigma _z}  P(z,w)=\int _{\Sigma _w} P(z,w)=0
\eea
In view of this property,  $\cB_2 ^{(0,1)} (\Omega)$ (defined in (\ref{3a3}))  is still invariant
under shifting the Green function by an arbitrary function $f$, namely $ G(z,w) \to G(z,w) + f(z) + f(w)$.

\subsection{The Arakelov Green function}

The Zhang--Kawazumi invariant $\varphi (\Omega)$ introduced in \cite{Zhang} and \cite{Kawazumi} 
is expressed in a number of equivalent forms which all involve the Arakelov Green function. 
The Arakelov Green function $\ln g(x,y)$ on $\Sigma \times \Sigma$ is symmetric 
$\ln g(x,y) = \ln g(y,x)$ and provides an inverse to the scalar Laplace operator on $\Sigma$,
just as the Green function $G$ of (\ref{2a9}) does, 
\bea
\label{3b1}
 \p  \, \bar \p \, \ln g(z,w) & = &  \pi \delta (z,w) -  \pi \mu _\Sigma (z)  
\no \\
 \p\,  \bar  \p \, G(z,w) & = & -2 \pi \delta (z,w) + 4 \pi \mu _\Sigma (z)  
\eea
where $\p = dz \p_z$ and $\bar \p = d\bar z \p_{\bar z}$ and $\int _\Sigma \delta (z,w)=1$ 
in local complex coordinates $z, \bar z$.  The normalization conditions  on $\mu _\Sigma$ are as follows,
\bea
\label{3b2}
\int _\Sigma \mu _\Sigma =1 \hskip 1in 
\int  _{\Sigma_z}  \mu _\Sigma (z) \ln g(z,w) =0
\eea
To define the form $\mu _\Sigma$, we proceed as follows. We shall keep the dependence on the  genus $h$
explicit whenever possible, though our main interest will be in the case $h=2$.
The canonical K\"ahler form $\mu$ on the Jacobian $J(\Sigma)$ of a  Riemann surface $\Sigma$
is defined by,
\bea
\label{3b3}
\mu = {i \over 2} \sum _{I,J} Y^{-1} _{IJ} d\zeta_I \wedge d\bar \zeta_J
\eea
The integrals of the holomorphic 1-forms $d\zeta_I$ along any closed cycle on $J(\Omega)$ 
are normalized to belong to $\ZZ^2 \oplus \Omega \ZZ^2$. Alternatively, in terms of a parametrization 
of $J(\Sigma)$ by real variables $x_I', x''_I \in \RR/\ZZ$, we have,\footnote{The notation with prime $x'$
and double prime $x''$ is borrowed from the representation of real characteristics, with which we shall soon 
identify these parameters.} 
\bea
\label{3b4}
\zeta_I = x''_I + \sum _J \Omega _{IJ} x'_J 
\hskip 1in 
\mu = \sum _I dx''_I \wedge dx'_I
\eea
The  Abel map  $j: \, z \to \zeta_I$ is defined by, 
\bea
\zeta_I (z) = \int ^z _{z_0} \omega_I - \Delta _I (z_0)
\eea 
where the Riemann vector is defined by
\bea
\label{riemanndef}
\Delta _I (z_0) = \half - \half \Omega _{II} + \sum _{J \not= I} \oint _{A_J} \omega_J (z) \int ^z _{z_0} \omega_I\,
\eea
The form $ \mu_\Sigma$ is defined as the pull-back under the Abel map $j$ of the 
canonical K\"ahler form $\mu$, divided by a factor of $h$ in order to achieve the 
normalization of (\ref{3b2}),
\bea
\label{3b5}
\mu _\Sigma (z) = {1 \over h} \, j _* \mu (z) = { i \over 2h} 
\sum _{I,J} Y^{-1} _{IJ} \omega_I (z) \wedge \overline{ \omega_J(z)}\,.
\eea
The Arakelov Green function $\ln g(z,w)$ is related to $G(z,w)$ by the shift,
\bea
\label{3b6}
\ln g(z,w) & = & - \half G(z,w) + f(z) + f(w) 
\no \\
f(z) & = & \half \int _\Sigma \mu _\Sigma (w) G(z,w) - \quart \int _{\Sigma^2} \mu _\Sigma (z) 
 G(z,w)  \mu _\Sigma (w)
\eea
Both integrals above are convergent, and $f(z)$ has been determined by enforcing the 
normalization condition (\ref{3b2}) on $\ln g$.

\subsection{The Zhang--Kawazumi invariant, $\varphi$}

For any genus $h$, the Zhang--Kawazumi invariant $\varphi (\Omega)$ of \cite{Zhang,Kawazumi} 
admits the representation,\footnote{In the mathematics literature, the 
Zhang-Kawazumi invariant $\varphi$ and the Faltings invariant $\delta$ are usually
denoted as functions of the surface, $\varphi (\Sigma)$ and $\delta (\Sigma)$ in order
to stress that they are real modular invariant functions of $\Omega$ and $\bar \Omega$
and thus depend only on the surface, not on the specific $\Omega$ chosen to represent $\Sigma$. 
Here we shall follows physics notation and denote both as functions of $\Omega$.}
\bea
\label{3c1}
\varphi (\Omega) =  \sum _\ell \sum _{I,J} { 2 \over \lambda _\ell} 
\left | \int _\Sigma \phi _\ell (z) \omega'_I(z) \wedge \overline{\omega'_J(z)} \right |^2
\eea
in a basis of Abelian differentials $\omega '$ normalized by 
$\int _\Sigma \omega _I' \wedge \overline{\omega '_J} =-2 i \delta _{IJ}$, and where 
$\lambda _\ell$ are the non-zero eigenvalues of the Laplace operator evaluated 
for the Arakelov metric on $\Sigma$, and $\phi _\ell$ are the corresponding eigenfunctions, 
normalized with respect to the volume form $\mu_\Sigma$. The Zhang--Kawazumi invariant 
$\varphi (\Omega)$ also admits the following equivalent representation \cite{Zhang},
\bea
\label{3cc1}
\varphi (\Omega)   =  \int _{\Sigma ^2} \nu (z,w) \, \ln g(z,w) \, .
\eea
where the bi-form $\nu (x,y)$ may be expressed as follows,\footnote{Note that the corresponding 
expression for $k$ in (2.5) and for $\nu$ in equation (2.6) of \cite{DeJong3} are incompatible with 
the normalization of the Abelian differentials implied by the pairing of (1.1). The problem may be 
traced to an inconsistent  change in normalization of the Abelian differentials effected in 
Proposition 2.5.3 of \cite{Zhang}. These inconsistencies have been resolved in 
writing our equation (\ref{3cc1}) and (\ref{3c2}).}
\bea
\label{3c2}
\nu _\Sigma (z,w) = 2 \mu _\Sigma (z) \wedge \mu _\Sigma (w)
+ \half \sum _{I,J,K,L} Y^{-1} _{IL} Y^{-1} _{JK} 
\omega _I(z) \wedge \overline{\omega_J(z)} \wedge \omega _K(w) \wedge \overline{\omega_L(w)} 
\eea
with the following normalization, 
\bea
\int _{\Sigma _z} \nu _\Sigma  (z,w) = (2-2h) \mu _\Sigma (w) 
\hskip 1in \int _{\Sigma ^2} \nu_\Sigma  (z,w) = 2-2h
\eea
Note that both representations of the Zhang--Kawazumi invariant are expressed in terms of the 
Arakelov Green function $\ln g$, and that neither formula is invariant under shifts 
$\ln g(z,w) \to \ln g(z,w) + f(z) + f(w)$ by an arbitrary function $f$.

\subsection{Proportionality of  $\varphi$ and $\cB _2 ^{(0,1)} $}

We will now show that the invariant $\varphi (\Omega)$, and the coefficient $\cB _2 ^{(0,1)} (\Omega) $ are 
simply proportional to one another. The first step in this proof uses the following relations between bi-forms,
which may be easily proven by inspection, 
\bea
\label{3d1}
P(z,w) = 2h \nu _\Sigma (z,w) + 4 h(h-1) \mu _\Sigma (z) \wedge \mu_\Sigma (w)
\eea
Next, we recast $\cB_2 ^{(0,1)}(\Omega) $ in terms of the Arakelov Green function in (\ref{3a3}), using the 
relation on the first line of (\ref{3b6}). The terms in $f$ cancel out in view of (\ref{3a5}), and we find, 
\bea
\label{3d2}
\cB _2 ^{(0,1)} (\Omega) =  16  \int _{\Sigma ^2}  P(z,w) \, \ln g (z,w)
\eea
Next, we express $P$ in terms of $\nu_\Sigma $ and $\mu _\Sigma $ using (\ref{3d1}), and make use of the 
defining relation of the Arakelov Green function in (\ref{3b2}) to drop the term in $\mu_\Sigma$. 
As a result, we find, 
\bea
\label{3d3}
\cB _2 ^{(0,1)} (\Omega) = 32h \varphi (\Omega)
\eea
An alternative way of stating the result is that the invariant $\varphi (\Omega)$ admits a simple
representation in terms of the Green function $G(z,w)$ by, 
\bea
\label{3d4}
\varphi (\Omega) = - { 1 \over 4h} \int _{\Sigma ^2} P(z,w) \, G(z,w)
\eea
This expression for $\varphi (\Omega)$ is now invariant under any shift 
$G(z,w) \to G(z,w) + f(z) + f(w)$.

\section{Higher-order invariants}
\setcounter{equation}{0}
\label{sec4}

A natural generalization of the Zhang-Kawazumi invariant $\varphi$ is obtained by 
considering higher order expansion terms of the superstring 4-point function, 
and more specifically of the unintegrated partial amplitudes $\cB _h (s,t,u | \Omega )$.

\sm

Recall that for genus $2$, we have $\cB _2 ^{(0,0)}=0$, while the coefficient $\cB_2 ^{(1,0)}$ 
is the constant which governs the $D^4 \cR^4$ correction.
Next, the coefficient $\cB _2 ^{(0,1)}$ produces the Zhang--Kawazumi invariant. Finally,
all coefficients $\cB_2^{(p,q)}$ with $p+q \geq 2$ produce new invariants which
generalize, in a way, the $\varphi$ invariant at genus two. The general form of the 
invariants $\cB_2^{(p,q)}$  is obtained by expanding the exponential to order
$n=2p+3q$ in all variables $s$, $t$, $u$, so that we have, 
\bea
\label{4b3}
\cB_2 (s,t,u | \Omega ) \bigg | _{n} \! \! & = & \! \!
{ 1 \over n!} \int_{\Sigma^4} { |\cY _S|^2 \over (\det \Im \Omega)^2} 
\bigg ( s\,G(1,2) +  t\,G(1,4) + u\,G(1,3) 
\no \\ && \hskip 1.3in 
 +  s\,G(3,4) +  t\,G(2,3)+ u\,G(2,4) \bigg )^n
\eea
Next, one recasts this homogeneous polynomial of degree $n+2$ into the symmetric functions 
$\sigma _2$ and $\sigma _3$. This combinatorial problem can be solved
with the help of a graphical expansion.

\subsection{The invariants $\cB_2^{(2,0)}$ and $\cB_2^{(1,1)}$}

In this section, we shall make the simplest of these generalizations as explicit as possible.
As examples, we shall work out in some detail the invariants of order low orders $(\sigma _2)^2$
and $\sigma _2 \sigma _3$. In view of the general analysis that leads to (\ref{4a3}), this contribution is proportional to $\sigma _2$,
a fact that may also be checked by direct calculation. To obtain the coefficient $\cB _2 ^{(2,0)} (\Omega)$
it will suffice to set $u=-t$ and $s=0$. To obtain $\cB _2 ^{(1,1)} (\Omega)$ one proceeds analogously, 
but sets $t=s$ and $u=-2s$ instead. One finds, 
\bea
\label{4b4}
\cB _2 ^{(2,0)} (\Omega) & = & 
\int _{\Sigma ^4} { | \Delta (1,2) \Delta (3,4) |^2 \over \det (\Im \Omega)^2}
\Big ( G(1,4) + G(2,3)  - G(1,3) - G(2,4) \Big )^2
\no \\
\cB _2 ^{(1,1)} (\Omega) & = & - { 1 \over 6^3}
\int _{\Sigma ^4} { | \Delta (1,2) \Delta (3,4) - \Delta (1,4) \Delta (2,3)|^2 \over \det (\Im \Omega)^2}
\Big ( G(1,2)+G(3,4)
\no \\ && \hskip 1.5in  + G(1,4)+G(2,3)  - 2G(1,3) - 2G(2,4) \Big )^3
\eea
These expressions are manifestly modular invariant, and convergent. They are also manifestly invariant
under shifting the scalar Green function $ G(z,w) \to G(z,w) + f(z) + f(w)$, so that the argument
may be expressed in terms of cross-ratios.

 \subsection{Diagrammatic expansion}
  
As in the case of the genus-one amplitude, the coefficients of the terms in the low energy 
expansion have an obvious graphical representation in terms of products of propagators.    
Since the amplitude has an overall measure that is of order $s^2\, \cR^4$, a diagram with 
$n$ propagators contributes to a term of order $s^{n+2}\, \cR^4$ that has a coefficient 
$\cB_2^{(p,q)}$, where $2p+3q=n+2$. 
An important qualitative difference between the genus-one and genus-two cases is that the zero mode part of the Green function does not decouple from the amplitude for genus $h>1$.  Consequently, there are non-zero contributions from diagrams in which one or more vertices are connected to a single propagator.   

The simplest example of a non-vanishing diagram with $h=2$ is  the single propagator, which gave zero contribution at genus one but contributes to $\cB_2^{(0,1)}$ (the integrand of the coefficient of $D^6\,\cR^4$), as discussed in this paper,
\vskip 0.6cm
\begin{center}
\centering
\begin{tikzpicture} [scale=1.5,line width=0.30mm]\scope[xshift=2.5cm]
\begin{scope}[xshift=1cm,yshift=1cm]
\draw (2,-0.5) node{$\bullet$} -- (3,-0.5)  node{$\bullet$} ;
\end{scope}
\endscope
\end{tikzpicture}
\label{singletwo}
\end{center}
\vskip 0.2cm

In the genus-one case there was only one diagram with two propagators that contributed 
to the expansion.   For  genus $h=2$ here are two additional diagrams that also contribute 
to $B_2^{(2,0)}$ (the integrand of the coefficient of $D^8\,\cR^4$).
\vskip 0.2cm

\begin{center}
\centering
\begin{tikzpicture} [scale=1.5,line width=0.30mm]
\scope[xshift=4.5cm]
\begin{scope}[xshift=1cm,yshift=1cm]
\draw (0,0) node{$\bullet$} ;
\draw (1,0) node{$\bullet$} ;
\draw (0, -1) node{$\bullet$} ;
\draw (1, -1) node{$\bullet$} ;
\draw (0,0) -- (1.0,0);
\draw (0,-1) -- (1,-1);
\draw(2,-0.5) ;
\end{scope}
\endscope
\begin{scope}[xshift=2cm]
\draw (0,0) node{$\bullet$} ;
\draw (0,1) node{$\bullet$} ;
\draw (0.7,0.5) node{$\bullet$} ;
\draw (0,0) -- (0.7,0.5);
\draw (0,1) -- (0.7,0.5);
\draw(1.6,0.5) ;
\end{scope}
\end{tikzpicture}
\label{singletwoA}
\end{center}

\vskip 0.2cm

In addition to the two diagrams with three propagators shown earlier for the genus-one case the following diagram contribute to the coefficient $B_2^{(1,1)}$  (the integrand of the coefficient of $D^{10}\,\cR^4$),

\vskip 0.2cm

\begin{center}
\centering
\begin{tikzpicture} [scale=1.5,line width=0.30mm]\scope[xshift=2.5cm]
\begin{scope}[xshift=1cm,yshift=1cm]
\draw (-0.5,-0.5) node{$\bullet$} ..controls (0,-0.8) ..  (.5,-0.5) node{$\bullet$};
\draw (-0.5,-0.5) node{$\bullet$} ..controls (0,-0.2) ..  (.5,-0.5) node{$\bullet$};
\draw (-0.5,-0.5) -- (-0.5,0.0)node{$\bullet$};
\end{scope}
\begin{scope}[xshift=1cm,yshift=1cm]
\draw (3,-0.5) node{$\bullet$} ..controls (3.5,-0.8) ..  (4,-0.5) node{$\bullet$};
\draw (3,-0.5) node{$\bullet$} ..controls (3.5,-0.2) ..  (4,-0.5) node{$\bullet$};
\draw (3, 0)node{$\bullet$} -- (4,0.0)node{$\bullet$};
\end{scope}
\endscope
\end{tikzpicture}
\label{tripletwo}
\end{center}

\vskip 0.2cm

\begin{center}
\begin{tikzpicture} [scale=1.5,line width=0.30mm]\scope[xshift=2.5cm]
\begin{scope}[xshift=0cm,yshift=1cm]
\draw (-0.5,0) node{$\bullet$} ;
\draw (0.5, 0) node{$\bullet$} ;
\draw (0.5, -0.5) node{$\bullet$} ;
\draw (-0.5, -0.5) node{$\bullet$} ;
\draw (0.5, -0.5) node{$\bullet$} ;
\draw (-0.5,0) -- (0.5,-0.5);
\draw (-0.5,-0.5) -- (0.5,-0.5);
\draw (0.5,0) -- (0.5,-0.5);
\draw(2,-0.5) node{$$};
\end{scope}
\begin{scope}[xshift=0cm,yshift=1cm]
\draw (4,0) node{$\bullet$} ;
\draw (3, 0) node{$\bullet$} ;
\draw (4, -0.5) node{$\bullet$} ;
\draw (3, -0.5) node{$\bullet$} ;
\draw (3, -0.5) node{$\bullet$} ;
\draw (4,0) -- (4,-0.5);
\draw (3,-0.5) -- (4,-0.5);
\draw (3,-0.5) -- (3,0);
\end{scope}
\endscope
\end{tikzpicture}
\label{tripletwoA}
\end{center}

At this order the following diagrams with more vertices contribute to the five-point and six-point functions,

\vskip 0.6cm

\begin{center}
\begin{tikzpicture} [scale=1.5,line width=0.30mm]\scope[xshift=2.5cm]
\begin{scope}[xshift=0cm,yshift=1cm]
\draw (-0.5,0) node{$\bullet$} ;
\draw (-0.5, -1) node{$\bullet$} ;
\draw (0.5, -1) node{$\bullet$} ;
\draw (-0.5, -0.5) node{$\bullet$} ;
\draw (0.5, -0.5) node{$\bullet$} ;
\draw (-0.5,0) -- (0.5,-0.5);
\draw (-0.5,-0.5) -- (0.5,-0.5);
\draw (-0.5,-1) -- (0.5,-1);
\end{scope}
\begin{scope}[xshift=0cm,yshift=1cm]
\draw (3,0) node{$\bullet$} ;
\draw (4,0) node{$\bullet$} ;
\draw (3, -1) node{$\bullet$} ;
\draw (4., -1) node{$\bullet$} ;
\draw (3, -0.5) node{$\bullet$} ;
\draw (4, -0.5) node{$\bullet$} ;
\draw (3,0) -- (4.,0);
\draw (3,-0.5) -- (4,-0.5);
\draw (3,-1) -- (4,-1);
\end{scope}
\endscope
\end{tikzpicture}
\label{highern}
\end{center}

\vskip 0.2cm

\section{Alternative forms and the Faltings invariant}
\setcounter{equation}{0}
\label{sec5}

The Zhang--Kawazumi invariant may be re-expressed in  a number of useful 
ways, of which perhaps the most important is via the Faltings $\delta$-invariant. 
It is not so much the Faltings invariant itself that is of use to us, but rather the circumstance 
that $\delta (\Omega)$ itself admits many alternative formulations. We shall not present 
a general definition of $\delta (\Omega)$ here, but rather we refer the interested reader to
\cite{Bost1, Bost2} for detailed information.

\sm

We begin by exhibiting the relation between the invariants $\varphi (\Omega) $ and $\delta (\Omega) $ 
obtained in Corollary 1.8 of \cite{DeJong2}, and specialized here to the case of 
genus two,\footnote{Note that the Faltings invariant, denoted  here and in 
 \cite{Bost1, Bost2} by $\delta$, is referred to as  $\delta _F$ in \cite{DeJong2}.}
\bea
\label{5a1}
 \varphi (\Omega) 
= 36 \ln 2 - 40 \ln ( 2 \pi) - 3 \ln \| \Psi _{10} (\Omega)  \| - {5\over 2} \delta (\Omega)
\eea
Here, $\Psi _{10} $ is the unique genus-two cusp modular form of weight 10 introduced by Igusa,
and $\| \Psi _{10}\|$ is its modular invariant Peterson norm, which are respectively defined by,
\bea
\label{5a2}
\Psi _{10} (\Omega) & = & \prod _{\delta \, {\rm even}} \tet [\delta ] (0, \Omega )^2
\no \\
\| \Psi _{10} (\Omega) \| & = & (\det Y)^5 |\Psi _{10} (\Omega) |
\eea
The genus-two $\tet$-function with general real characteristics $[x]$ is defined by
\bea
\label{5b7}
\tet   [ x ] ( \zeta , \Omega ) 
= \sum _{n \in \ZZ^2} \exp \Big \{ i \pi (n+x')^t \Omega (n+x') + 2 \pi i (n+x')^t (\zeta + x'') \Big \}
\eea
where the characteristics are parametrized in terms of $x'$ and $x''$ following (\ref{3b4}),
\bea
\label{5b6}
{} [x] = [x' ~ x''] 
\hskip 1in 
x' = \left [ \matrix{  x_1' \cr  x_2' \cr } \right ] \hskip 1in x'' = \left [ \matrix{  x''_1 \cr  x''_2 \cr } \right ]
\eea 
The $\tet$-function without characteristics is defined by $\tet (\zeta, \Omega) = \tet [0] (\zeta, \Omega)$.

\subsection{$\varphi$ as an integral over the Jacobian}

In \cite{Bost1, Bost2}, two alternative expressions are provided for the Faltings invariant 
$\delta (\Omega)$ at genus two. The first  is as an integral over the 
Jacobian $J(\Omega)$,\footnote{The integral of \cite{Bost1, Bost2} is originally 
to be carried out over ${\rm Pic}_1 (\Sigma)$,
the Picard variety of holomorphic line bundles over $\Sigma$ with first Chern class equal to 1.
Choosing an arbitrary reference point in ${\rm Pic}_1 (\Sigma)$, we use the 
standard isomorphism between ${\rm Pic}_1 (\Sigma)$ and $J(\Sigma)$ to recast the 
integral over $J (\Sigma)= J (\Omega)$.}
\bea
\label{5b1}
\delta (\Omega) = 12 \ln 2  - 16 \ln (2 \pi) - \ln \| \Psi _{10} (\Omega) \| 
-  \int _{J (\Omega)} \mu \wedge \mu \, \ln \| \tet \|^2
\eea
In this expression, $\mu$ is the canonical K\"ahler form on $J(\Omega)$ defined in (\ref{3b3}).
The Peterson norm of $\tet$ is defined for  $\zeta \in J(\Omega)$ as follows,
\bea
\label{5b2}
\| \tet \|^2  (\zeta , \Omega) = (\det Y)^\half |\tet (\zeta , \Omega)|^2 \exp \{ - 2 \pi (\Im \zeta )^t Y^{-1} (\Im \zeta ) \}
\eea
Expressed in terms of the integral over $J(\Omega)$, the Zhang--Kawazumi invariant takes the form,
\bea
\label{5b3}
 \varphi (\Omega) 
=  \varphi_0 - \half \ln \| \Psi _{10} (\Omega)  \| + \frac{5}{2} \int _{J (\Omega)} \mu \wedge \mu \, \ln || \tet ||^2
\eea
where $\varphi _0 = 6 \ln 2 $.
Remarkably, in this combined expression, the terms in $\ln ( \det Y)$ cancel one another, and the 
following simplified form may be obtained,
\bea
\label{5b4}
\varphi (\Omega) 
=   \varphi_0  - {1 \over 4}  \ln | \Psi _{10} (\Omega)  |^2 + 5  \ln  \Phi  (\Omega)\,,
\eea
where $\Phi  (\Omega)$ results from the integral over the Jacobian.  To represent this quantity, a particularly
convenient parametrization of $J(\Omega)$ is in terms of the real coordinates $x'_I$ and $x''_I$
introduced in (\ref{3b4}) and (\ref{5b6}). In terms of this parametrization,  we have,\footnote{We
thank Boris Pioline for pointing out an inconsistency of normalization, by a factor of 2 in the 
exponent of $\Phi$, in the first version of this paper.}
\bea
\label{5b5}
\Phi (\Omega) =
\exp \left \{  \int _{T^4} d^4x  \ln \Big | \tet  [ x  ] ( 0, \Omega ) \Big |^2  \right \} 
\eea
The measure of integration is $d^4 x = dx_1' dx_1'' dx_2 ' dx_2''$ which is subject to the 
relation $\mu \wedge \mu = 2 d^4x$, while the domain of integration is $T^4 = (\RR/\ZZ)^4$. 
In te passage from (\ref{5b3}) to (\ref{5b4}) we have also made use of the standard formula, 
\bea
\tet [x] (\zeta, \Omega) = \tet [0](\zeta + \Omega x' + x'', \Omega) 
\exp \Big \{ i \pi (x')^t \Omega x' + 2 \pi i (x')^t (\zeta + x'') \Big \}
\eea
Since $|\tet [x ](0,\Omega)|$ is invariant under shifts in $x$  by $\ZZ^4$, 
the range of integration $T^4$ may be replaced by $[0,1]^4$. For any fixed $\Omega$,
the integral over $x$ is  convergent. Thus, $ \Phi  (\Omega) $ is finite
throughout the interior of Siegel upper half space $H_2$, without poles or zeros. 
However, we shall see in section \ref{sec6} that $\Phi (\Omega)$ has singularities near the boundary
of moduli space.

\subsection{Modular properties}

Modular transformations  $M \in Sp(4, \ZZ)$ obey the defining relations,  
\bea
\label{5b10}
M = \left ( \matrix{ A & B \cr C & D \cr} \right )
\hskip 0.8in 
\cJ = \left ( \matrix{ 0  & -I  \cr I & 0 \cr} \right )
\hskip 0.8in 
M^t \cJ M = \cJ 
\eea
Their action on the period matrix in $H_2$ is given by, 
\bea
\label{5b11}
\Omega \to \tilde \Omega  = (A\Omega +B) (C \Omega +D)^{-1} 
\eea
while on real characteristics, we have, 
\bea
\label{5b12}
\left [ \matrix{ x' \cr x'' \cr} \right ] \to \left [ \matrix{ \tilde x' \cr \tilde x'' \cr} \right ] = 
\left ( \matrix{ D & -C \cr -B & A \cr} \right ) \left [ \matrix{ x' \cr x'' \cr} \right ]
+\half {\rm diag} \left ( \matrix{  CD^t \cr AB^t \cr} \right )
\eea
Their action on $\tet$-constants with characetristics takes the form, 
\bea
\label{5b13}
\tet [\tilde x ] (0 , \tilde \Omega ) 
= \det (C \Omega +D)^\half \tet  [x  ] (0, \Omega)
\eea
Thus, the modular transformation property of $\Phi $ is as follows,
\bea
\label{5b14}
\Phi  (\tilde \Omega) = \left |\det (C \Omega + D) \right | \, \Phi  (\Omega)
\eea
which makes it a real modular form of weight $(\half,\half)$.

\subsection{Holomorphy properties of $\Phi$}

Although formally we have $\ln |\tet|^2 = \ln \tet + \ln \bar \tet$, the form $\Phi$ is not the 
absolute value of a holomorphic modular form on moduli space. This property was shown in
\cite{Kawazumi} and \cite{DeJong1}. Here, we shall give an elementary derivation of this result,
and exhibit the difference in behavior between the genus-one and genus-two cases. 

\bigskip
\noindent {\bf $\bullet$ Genus two}
\sm

We proceed from (\ref{5b5}) by introducing an explicit  regulator $\ep$ for the logarithmic singularity of the integrand,
\bea
\label{5b15}
\Phi _\ep (\Omega) =
\exp \left \{ \int _{T^4} d^4x 
 \ln \left ( \Big  | \tet [x  ] ( 0, \Omega ) \Big |^2 + \ep ^2 \right )  \right \} 
\eea
In view of the integrability of the logarithmic singularity, we clearly have $\Phi _\ep \to \Phi$ as $\ep \to 0$.
We shall use  $\Phi_\ep$ to regularize the derivatives of $\Phi$, as usual.

\sm

We restrict attention to the variation along a single complex parameter $t$ in the Siegel upper 
half space $H_2$, with (locally) holomorphic dependence of the period matrix $\Omega _{IJ} (t)$ on~$t$.
Having already established that $\Phi_\ep $ is finite and non-zero everywhere on the interior of 
$H_2$, it suffices to compute the Laplacian in $t$ which is given by,
\bea
\p _t \p _{\bar t} \ln \Phi _\ep (\Omega (t))
=
\int _{T^4} d^4x \, \left | { \p \over \p t}  \tet [x] (0, \Omega (t)) \right |^2 \, 
{ \ep ^2 \over \left ( |\tet [x] (0,\Omega (t) )  |^2 + \ep^2 \right )^2}
\eea
As $\ep \to 0$, the integral over $T$ is supported on the subset of $J(\Omega)$ where $\tet$ vanishes,
\bea
\p _t \p _{\bar t} \ln \Phi (\Omega (t))
=
\int _{T^4} d^4x \, \left | { \p \over \p t}  \tet [x] (0, \Omega (t)) \right |^2 \, 
\delta ^{(2)} (\tet [x] (0, \Omega (t) )
\eea
The integrand is everywhere positive or zero, which makes the integral itself positive or zero.

\sm

There is an interesting geometrical interpretation of this formula in terms of the $\tet$-divisor,
which we shall denote by $\Theta $, and which is defined by,
\bea
\Theta (\Omega) = \{ \zeta \in J(\Omega) ~ \hbox{such that} ~ \tet (\zeta, \Omega)=0 \}
\eea
A variation in $t$ produces a variation 
$\delta _J \Theta$ in  $\Theta$ because the Jacobian changes with $\Omega (t)$.
But there is also another variation $\delta _x \Theta$ due to an  intrinsic co-moving change of $\Theta$. 
These contributions are most clearly disentangled by formulating the $\tet$-divisor without characteristics,
and parametrizing $\zeta$ by real characteristics, 
\bea
 \zeta _I = \Omega _{IJ} x_J' + x_I''
 \eea
As $t$ varies, $x', x''$ must vary, along with $\Omega$, to keep  $\zeta$ in the 
$\tet$-divisor, so that we must have,\footnote{We shall use the following notations: 
a dot refers to the derivative with respect to $t$; the derivatives with respect to $\Omega$ are 
denoted by $\p_{II} = \p /  \p \Omega _{II}$ and $\p_{IJ} = \half \p / \p \Omega _{IJ}$ when $J \not = I$;
and the derivatives with respect to $x_I' $ and $x_I''$ are denoted respectively by $\p_I '$ and $\p_I ''$.} 
\bea
( \dot \Omega _{IJ} x_J' + \Omega _{IJ} \dot x_J' + \dot x_I'') \p_I \tet (\zeta , \Omega  ) + \dot \Omega _{IJ} 
\p_{IJ} \tet (\zeta , \Omega ) =0
\eea
evaluated at $\Omega = \Omega (t)$. In the parentheses, the first term represents $\delta _J \Theta$, 
while the remaining two terms represent $\delta _x \Theta$. Equivalently,
in terms of $\tet$-functions with characteristics, the variation of $\Theta$ in $t$ is given by,
\bea
\dot x_I' \p' _I \tet [x](0,\Omega) +\dot x_I'' \p'' _I \tet [x](0,\Omega) + \dot \Omega _{IJ} \p_{IJ} \tet [x] (0, \Omega)=0
\eea
Combining this formula with the Laplace equation in $t$, we find, 
\bea
\p _t \p _{\bar t} \ln \Phi  
=  \int _{T^4} d^4x \, \Big | \dot x_I' \, \p' _I \tet [x](0,\Omega (t) ) +\dot x_I'' \, \p'' _I \tet [x](0,\Omega (t)) \Big |^2 ~
\delta ^{(2)} (\tet [x] (0, \Omega (t) )
\eea
The Laplacian in $t$ receives contributions from the intrinsic variation $\delta _x \Theta$ only. 
For genus two, this intrinsic variation is not everywhere vanishing, as $\Theta (\Omega)$ 
varies non-trivially in $J(\Omega)$ with $\Omega$.
The non-trivial variation of $\Theta$ with $t$ is illustrated in Figure 1. 
As a result, $\p_t \p _{\bar t} \ln \Phi \not= 0$, and the function $\ln \Phi $ is not pluri-harmonic. 
In \cite{Kawazumi} and \cite{DeJong1}, explicit  formulas for the Laplacian of $\varphi$ were derived in terms 
of characteristic classes.


\begin{figure}[htb]
\begin{center}
\includegraphics[width=3.8in]{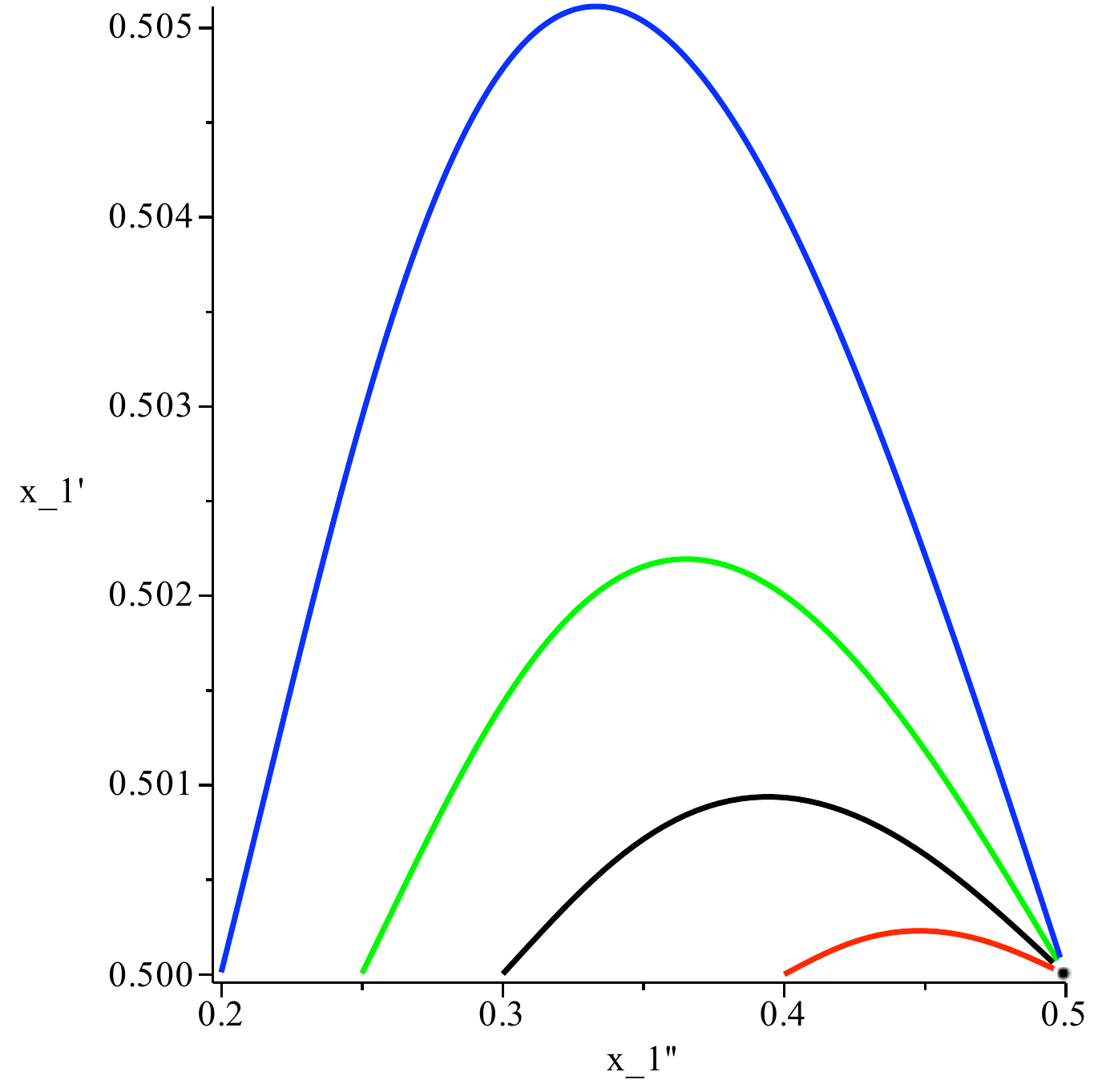}
\caption{Four one-dimensional slices of the genus-two $\tet$-divisor are presented
in co-moving coordinates $[x'x'']$ for the Jacobian.  The moduli $\Omega _{11}$,
$\Omega _{22}$, and the characteristic $x_2''$ are chosen ``generically":
we set $\Omega _{11}= 0.4+i$, $\Omega _{22}= 0.1+ 2i$, and $x_2''=0.55$. 
The remaining modulus is chosen to be real $\Omega _{12}=t$ and in the interval $[0,1]$.
At $t=0$, we choose $x_1'=x_1''=1/2$, a point which is on the $\tet$-divisor for any value of 
$x_2'$ in view of (\ref{6a1}). We plot the parametric  curves $x_1'(t)$ versus $x_1''(t)$  as $t$ runs 
from 0 to 1, for four values of $x_2'$, namely $x_2'=0.1$ (red), $x_2'=0.2$ (black), $x_2'=0.25$ (green),
and $x_2'=0.3$ (blue), such that $\tet [x] (0, \Omega)=0$.}
\end{center}
\label{fig:1}
\end{figure}


\bigskip
\noindent {\bf $\bullet$ Genus one}
\sm

However, the same arguments transposed to genus one lead to a different conclusion, as should 
have been expected from the explicit formula we have available for the genus-one Faltings invariant. The 
genus-one equivalent is given by, 
\bea
\p _\tau \p _{\bar \tau} \ln \Phi  (\tau)
=  \int _{T^2} d^2x \, \Big | \dot x' \, \p'  \tet [x](0,\tau ) +\dot x'' \, \p''  \tet [x](0,\tau) \Big |^2 ~
\delta ^{(2)} (\tet [x] (0, \tau )
\eea
For genus one, the $\tet$-divisor is a single point, $\zeta = 1/2 +\tau/2$, or in terms of 
characteristics $x' =x'' =1/2$. Thus, $\Theta$ has no intrinsic variation as $\tau$ is being varied, and hence 
we have, 
\bea
\p _\tau \p _{\bar \tau} \ln \Phi  (\tau)=0
\eea
This result is consistent with the explicit result $\Phi (\tau) = | \eta (\tau)|^2$.

\section{Issues involved in integrating $\varphi$ over moduli space}
\label{sec6}
\setcounter{equation}{0}

The coefficients of the terms in the low energy expansion of the string amplitude
at genus two are the integrated invariants, 
\bea
\int_{\cM_2} d\mu_2 \, \cB_2^{(p,q)}
\label{intinvs}
\eea
where $d\mu_2 =|d^3 \Omega |^2/ (\det \Im \Omega )^3$ is the $Sp(4,\ZZ)$-invariant measure
on the Siegel upper half space $H_2$.  
This suggests that it is of interest to consider the integral of $\varphi (\Omega)$ 
over the genus-$2$ moduli space.   Although we will not succeed in performing this integral explicitly we will 
prove that the integral is a well-defined, convergent expression.

\sm

To this end we recast $\varphi$ in another alternative form, 
\bea
\varphi (\Omega) = \varphi _0 + \half \sum _{\delta ~ {\rm even}}
\int _{T^4} d^4 x \, \ln \left | { \tet [x] (0, \Omega) \over \tet [\delta ] (0, \Omega)} \right |^2
\eea
To compute the integration over $\cM_2$, one would need to compute 
the following integral,
\bea
\Lambda [x] = \int _{\cM_2} d\mu_2\, \ln \left | { \tet [x] (0, \Omega) \over \tet [0 ] (0, \Omega)} \right |^2
\eea
$\Lambda [x]$ is periodic in each $x$ with period 1, and diverges when $x$ is an 
odd spin structure. In terms of $\Lambda [x]$, the integral of $\varphi$ over moduli
is given by,
\bea
\label{intofphi}
\int _{\cM_2} d\mu_2\,  \varphi 
= \varphi _0 \, V_2 + 5 \int _{T^4} d^4 x \Lambda [x] - \half \sum _{\delta ~ {\rm even}} \Lambda [\delta]
\eea
In order to prove that this integral is convergent we will analyse of the asymptotic properties 
of the integrand at the boundaries of moduli space where the genus-two surface degenerates, 
which  will be discussed next.

\subsection{Asymptotics of the $\varphi$ and $\delta$ invariants in degeneration limits}
\label{asymptotics}

In this subsection, we shall evaluate the limits of the invariants as the surface 
approaches the separating and non-separating degeneration nodes. These limits reproduce the 
genus-two results of \cite{Wentworth:1991},  where the degeneration limits of the Arakelov 
Green function and the Faltings invariant on a genus $h$ surface were considered, and the 
results of \cite{DeJong1} on the asymptotic  limits of $\varphi$.   Here we will start with the 
expressions for $\delta$ and  $\varphi$ in  (\ref{5b1}) and (\ref{5b3}), rewritten in terms 
of the modular form $\Phi $ which was defined in (\ref{5b5}),
\bea
\label{deltagain}
\delta & = & 12 \ln 2 -16 \ln (2 \pi) - 6 \ln (\det Y) - \ln \left | \Psi _{10} \right | - 2 \ln \Phi\,,
\\
\label{phiagain}
\varphi & = & 6 \ln 2  - \half \ln \left | \Psi _{10} \right | + 5 \ln \Phi\,.
\eea
To describe the degenerations we use the following parametrization of the period matrix,
\bea
\label{6a0}
\Omega = \left ( \matrix{ \tau _1 & \tau \cr \tau & \tau _2 \cr } \right )\,.
\eea
The separating degeneration is obtained by sending $\tau \to 0$ while keeping $\tau_1, \tau_2$
fixed. The non-separating degeneration is obtained by letting $\tau _2 \to i \infty$ 
while keeping $\tau_1, \tau$ fixed.  
Since the behavior of $\Psi_{10}$ at the degenerations is standard, we need only study 
the asymptotic behavior of $\Phi$ in the appropriate limits, details of which are given in 
appendix~\ref{degenerat}.

\bigskip
\noindent {\bf $\bullet$ Separating degeneration node $\tau \to 0$}
\sm

Substituting the expression for   $\Psi_{10}$ near the separating degeneration, 
\bea
\label{6a7}
\Psi _{10} (\Omega) = - 2^{12} (2 \pi \tau)^2 \eta (\tau_1)^{24} \eta (\tau_2)^{24} + \cO(\tau^4)\,,
\eea
together with the expression for $\Phi$  of (\ref{6a6}), into (\ref{phiagain})  gives the following limit for the Zhang--Kawazumi invariant $\varphi$,
\bea
\label{6a8}
\varphi (\Omega) =  - \ln \Big | 2 \pi \tau \eta (\tau_1)^2 \eta (\tau_2)^2 \Big | + \cO(\tau^2)
\eea
The Faltings invariant $\delta$ may now be derived from  (\ref{5a1}), and we find, 
\bea
\label{6a9}
\delta (\Omega) = \delta _1 (\tau_1) + \delta _2 (\tau_2)  
- \ln \Big | 2 \pi \tau \eta (\tau_1)^2 \eta (\tau_2)^2 \Big |^2 + \cO(\tau^2)
\eea
where the genus-one Faltings invariant on the degeneration component $I$ has been denoted by $\delta _I$,
and is given by (see for example \cite{DeJong3} after Proposition 4.6),
\bea
\label{6a10}
\delta _I (\tau_I) = - 8 \ln (2 \pi) - 6 \ln \Big ( ( \Im \tau _I  ) | \eta (\tau_I)|^4  \Big ) \,,
\eea
a combination which is manifestly modular invariant. The combination $2 \pi \tau \eta (\tau_1)^2 \eta (\tau_2)^2$, which is invariant under the $Sp(4,\ZZ)$  transformations (\ref{5b11}) to leading order in $\tau$, 
was identified in \cite{Wentworth:1991} as the intrinsic degeneration parameter (and referred to 
as $\tau$ in that paper).   With this identification, our final result (\ref{6a9}) for the separating degeneration 
precisely agrees with part $(a)$ of the Main Theorem of \cite{Wentworth:1991}, specialized to genus $h=2$ and $h_1=h_2=1$.

\bigskip
\noindent {\bf $\bullet$ Non-separating degeneration node $\tau_2 \to i \infty$}
\sm

Using the non-separating degeneration limit of $\Psi_{10}$, 
\bea
\label{psiten}
\Psi _{10} (\Omega) = - 2^{12} e^{2 \pi i \tau_2 } \eta (\tau_1)^{18} \tet _1 (\tau, \tau_1)^2
\eea
and the asymptotics for $\Phi$ of (\ref{6b6}), we find, 
\bea
\label{phidelone}
\varphi (\Omega) & = &  {\pi \over 6} (\Im \tau _2) 
+ { 5 \pi \over 6} { (\Im \tau)^2 \over \Im \tau_1}
- \ln \left | { \tet _1 (\tau, \tau_1) \over \eta (\tau_1)} \right |
\\
\delta (\Omega) & = & \delta _1 (\tau_1) - 8 \ln (2 \pi) + {7 \pi \over 3} \Im \tau_2 
- 6 \ln (\Im \tau_2) - { \pi \over 3} { (\Im \tau)^2 \over \Im \tau_1} 
- 2 \ln \left | { \tet _1 (\tau, \tau_1) \over \eta (\tau_1)} \right |
\no
\eea
Throughout this subsection we shall neglect contributions which vanish as $\tau_2 \to i \infty$.
Expressed in terms of the Arakelov Green function $\ln g(z)= \ln g(z|\tau_1)$
for modulus $\tau_1$ presented in (\ref{arakelovnorm}), these invariants become, 
\bea
\label{phidelarak}
\varphi (\Omega) & = &  {\pi \over 6} \left ( \Im \tau _2 - { (\Im \tau)^2 \over \Im \tau_1} \right )
- \ln g (\tau|\tau_1)
\\
\delta (\Omega) & = & \delta _1(\tau_1) - 8 \ln (2 \pi) + {7 \pi \over 3} \left ( \Im \tau_2 -  { (\Im \tau)^2 \over \Im \tau_1} \right )
- 6 \ln (\Im \tau_2)  - 2 \ln g(\tau|\tau_1) 
\no
\eea
In terms of the modular invariant degeneration parameter $|t|$ which was introduced in \cite{Wentworth:1991}, 
and is defined by,
\bea
\label{tauredef}
\Im \tau _2 -  { (\Im \tau)^2 \over \Im \tau_1} = - { 1 \over 2 \pi} \ln |t| + {1 \over \pi } \ln g (\tau|\tau_1)
\eea
the invariants take the following form,
\bea
\label{phidel}
\varphi (\Omega) & = & - {1 \over 12} \ln |t|   - {5 \over 6} \ln g (\tau|\tau_1) 
\\
\delta (\Omega) & = & \delta _1 (\tau_1)  -{7 \over 6} \ln |t| - 6 \ln ( - \ln |t|) +{1 \over 3} \ln g(\tau|\tau_1) - 2 \ln (2\pi) 
\no
\eea
This expression agrees precisely with part $(b)$ of the Main Theorem of \cite{Wentworth:1991} for genus two.

\subsection{Convergence of the integral over moduli space}
\label{convergence}

The preceding analysis of asymptotic behavior enables us to prove the 
convergence of the integral of $ \varphi$ over the genus-two moduli space $\cM_2$
with the measure $d\mu_2$, encountered  in (\ref{intofphi}).  The function $\varphi (\Omega)$ 
is well-defined everywhere in the interior of $\cM_2$, but has singularities as one approaches
the boundary of $\cM_2$. To deal with the boundary behavior in a systematic way, it 
will be convenient to replace $\cM_2$ by its Deligne-Mumford \cite{DM} compactification $\overline{ \cM_2}$,
which is obtained from $\cM_2$ by adjoining the divisors (a divisor is a subvariety of
complex co-dimension 1) corresponding to the separating node and to the non-separating node.
The integration measure $d\mu_2$ extends to a finite measure on $\overline{\cM_2}$
with finite volume. 

\sm

To show convergence of the integral of $\varphi$ on the compact space $\overline{\cM_2}$, 
it will suffice to show that the integral converges near each one of the compactification divisors. 
The divisors intersect, but the convergence of the integral near the intersection will be shown to follow
from the convergence near each divisor separately. 

\sm

$\bullet$ The asymptotic behavior of the measure near the separating divisor $\tau \to 0$  is given by, 
\bea
\label{sepmeas}
d\mu_2  \to 
|d^2\tau|\, \frac{|d^2\tau_1|}{(\Im \tau_1)^3} \,\frac{|d^2\tau_2|}{(\Im \tau_2)^3}\left(1+ \cO(|\tau|^2)\right)\,.
\eea
Since the most singular term as $\tau\to 0$  is given from (\ref{6a8}) by 
$\varphi \sim -\ln |\tau| + \ldots$ the $\tau$ integral converges near $\tau \to 0$, in view of 
the integration range of (\ref{domain}).

\sm

$\bullet$ The asymptotic behavior of the measure near the non-separating divisor $\tau_2 \to i\infty$ is 
similarly given by the following formula,
\bea
\label{nonsepmeas}
d\mu_2  \to |d^2\tau|\, 
\frac{|d^2\tau_1|}{(\Im \tau_1)^3} \,\frac{|d^2\tau_2|}{(\Im \tau_2)^3}\left(1+ \cO(\Im \tau_2^{-1})\right)\,.
\eea
From (\ref{phidelone}) we see that $\varphi \sim \pi \, \Im \tau_2/6+ \cO(\tau_2^0)$ 
so that the $\tau_2$ integral converges, in view of the integration range of (\ref{domain}).

\sm

$\bullet$ The asymptotic behavior near the intersection of the separating and non-separating divisors
is given by either formula (\ref{sepmeas}) or (\ref{nonsepmeas}) for the measure. The asymptotic 
behavior of $\varphi$ near the intersection of the divisors may be obtained either as the $\tau_2 \to i \infty$ asymptotics of (\ref{6a8}), or as the $\tau \to 0$ asymptotics of (\ref{phidelarak}). Happily, these
two limits are interchangeable, and give rise to the following uniform asymptotics near the 
intersection of divisors, 
\bea
\varphi (\Omega) = - \ln (2 \pi) + { \pi \over 6} \Im \tau_2 - \ln |\tau| - \ln |\eta (\tau_1)|^2
\eea
up to terms that vanish as $\tau \to 0$ and $\tau_2 \to i \infty$. It is readily seen that the 
convergence near the intersection is automatic once the convergence near each 
divisor has been checked. 

We conclude that the integral with the measure $d \mu_2$ of $\varphi$ over the compactified 
moduli space $\overline{\cM_2}$, and thus over the moduli space $\cM_2$, is convergent.

\section{Value of integrated invariant from $SL(2,\ZZ)$-duality} 
\label{sec7}
\setcounter{equation}{0}

Although we have not evaluated the integrated invariant directly, we will now determine 
the relationship of its value to the coefficient of the genus-two $D^6\, \cR^4$ term in the 
low energy expansion of the four-string amplitude in Type IIB string theory..

\sm

As we described in the introduction, the Type IIB theory is  invariant 
under the duality group $SL(2,\ZZ)$, which acts on the complex coupling  $T  = T _1+iT _2 $ 
by M\"obius transformations. The $SL(2,\ZZ)$ transformation properties of the other fields 
will not concern us here.   

\sm

Symmetry of the amplitude under the interchange of external states again implies 
that the low energy expansion of the analytic part of the  amplitude is a symmetric 
function of powers of the Mandelstam variables and has an expansion in powers of 
$\sigma_2$ and $\sigma_3$ in which each term is  invariant under $SL(2,\ZZ)$.  
These conditions imply that the analytic part of the full 
(i.e., non-perturbative)  amplitude has a low energy expansion of the form,
\bea
{\cal A}^{(4)}(\epsilon_i,k_i,T ) \Big|_{{\rm an.}} 
& = &   \kappa_{10}^2 \, {\cal R}^4  \Big ( T _2^2 \frac{3}{\sigma_3}
+ T _2^{\frac{1}{2}}\,{\cal E}_{(0,0)}(T )\,   + 
 T _2^{-\frac{1}{2}}\, {\cal E}_{(1,0)}(T ) \, \sigma_2 
 \no \\ && \hskip 1in 
 + T _2^{-1}\, {\cal E}_{(0,1)} (T )\, \sigma_3+ \ldots  \Big ) \,,
 \label{symmamp}
\eea
where the explicit powers of $T _2$  disappear after transforming from the 
string frame to the Einstein frame (in which the curvature, ${\cal R}$ is inert under $SL(2,\ZZ)$).  
The coefficients ${\cal E}_{(p,q)}(T )$ are  $SL(2,\ZZ)$-invariant functions.   
The prefactor of ${\cal R}^4$, which arose in the perturbative examples discussed earlier,  
in fact multiplies the full amplitude as can be deduced from maximal supersymmetry.  
The first term in the above expansion is the lowest order term in the tree-level expansion, 
which is equal to the tree-level supergravity amplitude.    The challenge is to determine 
the modular invariant coefficient functions of the higher order terms.  These are functions 
of $T $ and their expansions in the weak-coupling limit $T_2 \to \infty$ should start with 
power-behaved terms that correspond to terms in string  perturbation theory.

\sm

The first term in the $\alpha'$ expansion (\ref{symmamp}) beyond the supergravity amplitude is  
of order $\cR^4$, and corresponds to an  interaction which preserves 16 supersymmetries in an 
effective action, that may be expressed as an  integral over 16 Grassmann coordinates.
The next term is of order $D^4 \cR^4$, which is associated with an effective interaction which 
preserves 8 supersymmetries,  that may be expressed as an integral over 24 superspace Grassmann
coordinates.  These have $T $-dependent coefficients \cite{Green:1997tv, Green:1998by, Sinha:2002zr}
\bea
{\cal E}_{(0,0)}(T ) =2\zeta(3)\, E_{\threeh}(T )\,,\qquad {\cal E}_{(1,0)}(T ) = \zeta(5)\, \,E_{\fiveh}(T )\,,
\label{rfour}
\eea  
where $E_s(T )$ is an $SL(2,\ZZ)$ non-holomorphic Eisenstein series, which was 
encountered earlier in a different context and was defined in (\ref{dtwodef}).  
Although these solutions were initially discovered by indirect means they were 
subsequently determined by supersymmetry, which constrains the coefficients to 
satisfy  Laplace eigenvalue equations of the form  (\ref{laplaceeigenvalue}) with 
$s=3/2$ (in the $\cR^4$) case  or $5/2$  (in the $D^4\cR^4$ case)   \cite{ Green:1998by, Sinha:2002zr}.   

\sm

The  perturbative and non-perturbative content of these coefficients can easily be 
extracted by considering the Fourier modes of  $E_s(T )$, defined by
 \bea
2\zeta(2s)\,E_s(T ) = \sum_{N\ne 0} {\cal F}_N(T _2)\, e^{2i\pi NT _1}\,.
\label{eisendef3}
 \eea
 The non-zero modes ${\cal F}_{N\neq 0}(T _2)$ contain the effects of D-instantons, 
 with exponentially suppressed asymptotic behavior, ${\cal F}_N(T _2)\sim e^{-2\pi|N|T_2}$, 
 at weak coupling ($T _2 \to \infty$). The zero mode, on the other hand, is a sum of 
 two power behaved terms $T _2^{s}$ and $T _2^{1-s}$ which 
 correspond to particular terms in string perturbation theory,
 \bea
 {\cal F}_0(T _2) =  2\zeta(2s)\, T _2^s  +  \frac{ 2 \pi^{1/2}  
 \Gamma(s-1/2)}{\Gamma(s)} \,\zeta(2s-1) \, T _2^{1-s}\,,
 \label{zeromode}
 \eea
 Substituting the zero mode parts of the coefficients ${\cal E}_{(0,0)}(T )$ and 
 ${\cal E}_{(1,0)}(T )$, as defined in (\ref{rfour}),  into (\ref{symmamp}) gives the 
 contributions that are power-behaved in the coupling constant, 
 Thus, the perturbative contribution  to the $\cR^4$ term is obtained by setting $s=3/2$,
  \bea
 T_2^{\half}\, \int_{-\half}^\half  {\cal E}_{(0,0)}(T) \, dT_1
 =2\zeta(3)\,T_2^{\half}\, \int_{-\half}^\half E_{\threeh}(T)  \,  dT_1
 =  2\zeta(3) T_2^2 + 4 \zeta(2) \,,
\label{r4}
  \eea
which contains the sum of tree-level and one-loop contributions.   
Similarly, the perturbative contribution to the coefficient of $\sigma_2\, \cR^4$ is 
  \bea
T_2^{-\half}\, \int_{-\half}^\half  {\cal E}_{(1,0)}(T) \, 
dT_1=\zeta(5)\, T_2^{-\half}\, \int_{-\half}^\half E_{\fiveh}(T)  \, 
dT_1=  \zeta(5) T_2^2 + {4\over 3} \zeta(4) T_2^{-2}\,,
\label{d4r4}
  \eea
which contains the sum of tree-level and two-loop contributions.
The precise coefficients of the perturbative terms in (\ref{r4}) and (\ref{d4r4}) match those 
determined directly  from perturbative string calculations reviewed in section~\ref{zeroone}.  
The tree-level terms were shown in (\ref{treeexpand}) while the  genus-one term in  
(\ref{r4}) is given (up to a normalization factor) by $\int d\mu_1\,\cB_1^{(0,0)}$.  
Similarly, the genus-two term in (\ref{d4r4}) is given (up to a normalization factor)  
by $\int d\mu_2 \, \cB_2^{(1,0)}$  \cite{D'Hoker:2005ht}.
 This also accounts for the absence of a one-loop contribution to $\sigma_2 {\cal R}^4$ 
 in the ten-dimensional theory.
Generalizations of these results to lower dimensional theories with maximal 
 supersymmetry obtained by toroidal compactification involve combinations of 
 Eisenstein series for higher-rank duality groups, which are functions of more 
 moduli  \cite{Obers:1999um,Green:2010wi, Green:2010kv} (see also \cite{Pioline:2010kb}).

\sm

The coefficient of the  term $D^6 \cR^4$ in the low energy expansion which 
preserves 4 supersymmetries, ${\cal E}_{0,1}$,  is not an Eisenstein series but is 
expected to be a solution of the inhomogeneous Laplace equation
\bea
(\Delta_T   - 12)\, {\cal E}_{(0,1)}(T ) = - (2\zeta(3)\, E_\threeh)^2\,
\label{inhomo}
\eea
This equation was motivated by M-theory considerations in \cite{Green:2005ba} based 
on considering the compactification of Feynman diagrams of eleven-dimensional 
supergravity on a torus.  The solution to this equation has an asymptotic expansion for 
large $T _2$  that gives a contribution to the coefficient of the $\sigma_3 \,\cR^4$ term 
in (\ref{symmamp})  of the form
\bea
\frac{1}{T_2}\int_{-\half}^\half {\cal E}_{(0,1)}(T )\, dT_1 = \frac{2}{3} \zeta(3)^2 T _2^2 +\frac{4}{3}  \zeta(2)\zeta(3)  + {8\over 5}\zeta(2)^2  T _2^{-2} +{4\over 27}\zeta(6) T _2^{-4} + \cO(e^{-2\pi T _2})\, 
\label{ethreehthreeh}
\eea
which contains four perturbative terms that are power-behaved in $T _2$ that 
correspond to tree-level, one-loop, two-loop and three-loop string theory
contributions,  together with an infinite sum of  D-instanton contributions.  
The ratio of the tree-level and one-loop contributions agrees with the explicit 
string perturbation theory calculations and the overall normalization has been 
chosen to be consistent with a tree-level amplitude normalized to $3/\sigma_3$.

\sm

We can now compare the ratio of the two-loop perturbative contribution to 
${\cal E}_{(0,1)}\,\sigma_3\, \cR^4$ with the two-loop contribution to  
${\cal E}_{(1,0)}\, \sigma_2 \, \cR^4 $.
First note that the expressions for ${\cal E}_{(1,0)}$ and  ${\cal E}_{(0,1)}$ in 
(\ref{d4r4}) and (\ref{ethreehthreeh}) have been normalized to ensure that their 
tree-level contributions  have the correct relative normalizations, which accords 
with the tree-level expansion of the amplitude as given in  (\ref{treeexpand}),
\bea
 \cA ^{(0)} _4 \Big |_{\sigma_2\, \cR^4 + \sigma_3\, \cR^4} 
 = \kappa _{10}^2 T_2^2  \left ( \zeta(5) \,\sigma_2  + {2\over 3} \zeta(3)^2\, \sigma_3 \right ) \cR^4 .
 \label{treesum}
 \eea
 The ratio of two-loop contributions to the $\sigma_2\, \cR^4$ and $\sigma_3\, \cR^4$ 
 terms  (the $T_2^{-2}$ terms in  (\ref{d4r4}) and (\ref{ethreehthreeh})) is given by
\bea
 \frac{ \int_{\cM_2} d\mu_2\, \cB_2^{(0,1)}}{\int_{\cM_2} d\mu_2\,\cB_2^{(1,0)}} = {\frac{8}{5}\zeta(2)^2\over \frac{4}{3}\zeta(4)}= 3\,.
\label{ratiores}
\eea
From (\ref{3d3})  with $h=2$ we see that this means that the integral of the  
Zhang--Kawazumi invariant should take the value
\bea
\label{predict}
\int_{\cM_2} d\mu_2 \, \varphi 
= \frac{1}{64} \int_{\cM_2} d\mu_2 \, \cB_2^{(0,1)} 
=  \frac{3}{64} \,\int_{\cM_2} d\mu_2 \, \cB_2^{(1,0)}
=  \frac{3}{2}  V_2 =  \frac{2 \pi^3}{45} \,,
\eea
 where $V_2=4\pi^3/135$ is the volume of $\cM_2$ and we have substituted the 
 value $\cB_2^{(1,0)}=32$  obtained in (\ref{2c2}).  We note here that, by the 
 construction of $\varphi$ given in (\ref{3c1}), we have $\varphi (\Omega) >0$
 for all $\Omega$, which is consistent with the sign of the proposed relation 
 (\ref{predict}).

\sm

 It would be satisfying to find a method of evaluating $\int_{\cM_2} d\mu_2 \, \varphi$ directly, 
 which would provide a precise check on the $SL(2,\ZZ)$-duality prediction and might point to some 
 interesting mathematical properties of $\varphi$.

\vskip 0.1in 
 
\Acknowledgement

\sm

We thank Richard Wentworth for useful discussions. We also thank Boris Pioline for pointing 
out an important discrepancy of normalization, by  a factor of 2,  in the first version of this paper.
We are grateful for the support of the National Science Foundation under Grant 
No. PHYS-1066293 and the hospitality of the Aspen Center for Physics during the Summer 
of 2011 when this work was started and  the Summer of 2013 when it was completed.  
The research of ED is supported in part by NSF grant PHY-07-57702.
MBG  also acknowledges funding from the European Research Council under the European 
Community's Seventh Framework Programme (FP7/2007-2013) / ERC grant agreement no. [247252].


\appendix

\section{ Expressing $\varphi$ as a single integral over $\Sigma$}
\label{newform}
\setcounter{equation}{0}

For completeness, we here determine a 
second alternative expression for $\varphi$ based on \cite{Bost1, Bost2} that is given in terms of a single integral
over the surface $\Sigma$. To obtain this expression, we start from the following expression 
for the Faltings invariant, given in \cite{Bost1}, 
\bea
\delta (\Omega) = - 12 \ln (2 \pi) - 2 \ln \det Y - 2 \ln | \cM_{\nu_p \nu_q} | +  \int _{\Theta + p -q} \mu \ln \| \tet \|^2
\eea
In this formula, $p,q$ are  two distinct branch points, and $\nu_p$ and $\nu_q$ their associated 
odd spin structures. The expression for $\delta (\Omega)$ is independent of the choice made for $p,q$.
The modular object $\cM_{\nu_p \nu_q}$ entered the calculations of the genus-two 
superstring measure in \cite{D'Hoker:2001qp}, and may be expressed with the help of $\tet$-constants 
and of the six distinct genus-two odd  spin structures $\nu_i$ with $i=1,\cdots, 6$,
\bea
\cM_{\nu_i \nu_j} ^2 = \pi ^4 \prod _{k\not= i,j} \tet [\nu _i+ \nu_j + \nu_k] (0, \Omega )^4
\eea
Finally,  $\Theta$ is the $\tet$-divisor, namely the set of points  $\zeta \in J(\Omega)$ 
such that $\tet (\zeta, \Omega)=0$.
Using the Riemann vanishing theorem in genus-two, $\Theta$ may be parametrized as follows,
\bea
\Theta = \left \{ \zeta_I = \int ^z _{z_0} \omega_I - \Delta _I (z_0), ~~ z \in \Sigma \right \}
\eea
where $\Delta _I (z_0)$ is the Riemann vector defined in(\ref{riemanndef}).
Next, we proceed to reformulate the integral over the shifted $\tet$-divisor in terms of more familiar objects.
To do so, we use the relation $\nu _p = p - \Delta$ to recast the $\tet$-function into one with spin structure 
characteristic $ \nu_p$, 
\bea
\label{5a18}
\ln \| \tet (z +p-q-\Delta, \Omega) \|^2 & = & \ln | \tet [\nu_p] (z-q, \Omega) |^2
+ \half \ln (\det Y) 
\no \\ && - 2 \pi \sum _{I,J} Y^{-1}_{IJ} \left ( \Im \int ^z_q \omega_I \right ) \left (  \Im \int ^z_q \omega_J \right )
\eea
We make use of the prime form expressed with respect to spin structure $\nu_p$, 
\bea
\label{5a19}
E(z,w) = { \tet [\nu_p]  (z-w, \Omega) \over h_{\nu_p} (z)\,  h_{\nu_p} (w)}
\eea
for $w=q$, where $h_\nu(z)$ is the normalized holomorphic 1/2 form with odd spin structure $\nu$.
Recasting the first term in (\ref{5a18}) in terms of the prime form, we find,   
\bea
\label{5a20}
\ln \| \tet (z + p-q -\Delta, \Omega) \|^2 & = & \ln | E(z,q) |^2 + \ln | h_{\nu_p} (z) |^2 + \ln | h_{\nu_p}  (q)|^2 
\no \\ &&
+ \half \ln (\det Y)  - 2 \pi  \Im (z-q) ^t Y^{-1} \Im (z-q) 
\eea 
Combining the first and the last terms, we recognize the appearance of $- G(z,q)$, where the scalar Green function
$G$ was defined in (\ref{2a9}). Putting all together, $\varphi$ takes the form, 
\bea
\label{5a21}
\varphi (\Omega) & = & \tilde \varphi _0 - {1 \over 2} \ln |\Psi _{10} | 
-5 \ln \left | { h_{\nu_p}  (q)^2 \over \cM_{\nu_p, \nu_q}} \right |  
+ 5 \int _\Sigma \mu_\Sigma (z)  \left \{ G(z,q) - \ln | h_{\nu_p} (z) |^2 \right \}
\eea
Note that the combination $h_{\nu_p}  (q)^2 / \cM_{\nu_p, \nu_q} $ is independent 
of the branch point $\nu_p$. Given that the total expression for $\varphi$ is independent
of the branch points $p,q$, we see that the integral in (\ref{5a21})
must be independent of the branch point $p$.

\newpage

\section{Derivation of asymptotic limits of $\Phi$}
\label{degenerat}
\setcounter{equation}{0}

In this appendix we derive the asymptotic limits quoted in section~\ref{asymptotics}.

\subsection{The separating degeneration}

To leading order in the separating degeneration $\tau \to 0$, the genus-two $\tet$-function 
tends to,
\bea
\label{6a1}
\tet \left [ \matrix{ x_1 ' & x_1 '' \cr x_2 ' & x_2'' \cr} \right ] (0, \Omega )
= \tet [x_1 ' ~ x_1''] (0, \tau_1) \, \tet [x_2 ' ~ x_2''] (0, \tau_2) + \cO (\tau^2)
\eea
where $\tet [x_I ' ~ x_I''] (0, \tau_I) $ are the genus-one $\tet$-functions with 
real characteristics $x_I' ~ x_I ''$, defined in (\ref{2a11}). They may be expressed in terms of the
$\tet$-function with zero characteristics by, 
\bea
\label{6a3}
\tet [x_I' ~x_I''](0,\tau_I) = e^{ \pi i \tau_I (x_I')^2 + 2 \pi i x_I'x_I'' } \tet (x_I' \tau_I + x_I'', \tau_I)
\eea
The limit of $\Phi$ then becomes, 
\bea
\label{6a2}
\ln \Phi = \prod _{I=1}^2 \left ( \int _0 ^1 dx_I' \int _0 ^1 dx_I '' \, 
\ln | \tet [x_I ' ~ x_I''] (0, \tau_I) |^2 \right ) + \cO (\tau^2)
\eea
To evaluate each factored integral, we first express the $\tet$-function with characteristics 
in terms of $\tet$ with zero characteristics using (\ref{6a3}), 
and then use the infinite product representation of the latter, 
\bea
\label{6a4}
\ln \tet (z, \tau) = \sum_{n=1}^\infty  \ln \left[ \left ( 1 - e^{2 \pi i n \tau} \right )
  \left ( 1 - e^{\pi i ( 2 n \tau + 2 z -1 -\tau)} \right ) \left ( 1 - e^{\pi i ( 2 n \tau - 2 z -1 -\tau)} \right )\right]\,.
\eea
The $x'$ and $x''$ integrals in (\ref{6a2})  may be carried out by considering each of the three factors in the square parentheses in turn, as follows.  The first factor trivially gives
\bea
\label{intone}
\int _0 ^1 \! dx'_I \int _0 ^1 \! dx''_I \, \sum_{n=1}^\infty \ln\left ( 1 - e^{2 \pi i n \tau} \right ) =\sum_{n=1}^\infty \ln\left ( 1 - e^{2 \pi i n \tau} \right )\,.
\eea    
The second factor contributes zero, since the  logarithm  has a Taylor expansion  in powers 
of the exponential in its argument and the modulus of the exponential is strictly less than $1$.  
Each term in the Taylor series vanishes by virtue of the $x''$ integral.  
In the third factor, terms with $n>1$ vanish analogously, but for the $n=1$ term the expansion breaks 
down when $\half \le x'\le 1$. Thus, that region needs to be re-expanded by writing the 
argument of the logarithm as $( 1 - e^{\pi i (\tau-x'\tau -2x'' -1)}) 
=  -e^{\pi i (\tau - 2x'\tau -2x'' -1)}(1- e^{-\pi i (\tau - 2x'\tau -2x'' -1)})$ 
leading to a non-zero contribution given by 
\bea
\label{lndetail}
\int _\half ^1 \! dx'_I \int _0 ^1 \! dx''_I \, i \pi \left((1 -2x')\tau - 2x''-1\right) +i\pi = -\quart i \pi \tau  \,.   
\eea
In addition to the above terms, in converting from $\ln \tet (z,\tau)$ to $\log \tet [x',x''](0,\tau)$ 
using (\ref{6a3}) we also need to evaluate the logarithm of the prefactor 
\bea
\label{expint}
 \pi i\int _0 ^1 \! dx' \int _0 ^1 \! dx'' \,( \tau (x')^2 + 2  x'x'') =  \pi i\int _0 ^1 \! dx'  \,( \tau (x')^2 + x') = \frac{1}{3} i\pi \tau + \half i\pi 
\eea
Combining (\ref{intone}), (\ref{lndetail}), and (\ref{expint}), the final result is
\bea
\label{6a5}
\int _0 ^1 \! dx'_I \int _0 ^1 \! dx''_I \,  \ln | \tet [x_I'  ~ x_I''] (0, \tau_I) |^2
= \ln \left | \eta (\tau_I) \right |^2
\no
\eea
As a result, the asymptotics of $\Phi$ is given by
\bea
\label{6a6}
\Phi =  | \eta (\tau_1)  \eta (\tau_2)|^2 + \cO(\tau^2)\,.
\eea

\subsection{Non-separating degeneration}

In terms of the parametrization (\ref{6a0}) the non-separating degeneration is given by
$\tau_2 \to i \infty$. Later on, we shall be more precise in the finite part of this limit.
To extract the leading asymptotics of the genus-two $\tet$-function, given by (\ref{5b7}), 
we isolate the $\tau_2$-dependence by recasting the double sum over $n_1,n_2$ as a 
simple sum over $n=n_2$ of genus-one $\tet$-functions.
For brevity, we shall use the notation $x_I = x_I'$ and $y_I = x_I''$. We find, 
\bea
\label{6b1}
\tet [x] (0, \Omega) & = & \sum _{n\in \ZZ} C_n \, \tet (x_1 \tau _1 + (n+x_2)\tau +y_1 , \tau_1)
\\
C_n  & = & 
\exp i \pi \Big \{ \tau_1 x_1^2 + \tau _2 (n+x_2)^2 + 2 \tau x_1 (n+x_2) + 2 x_1 y_1 + 2 (n+x_2) y_2 \Big \}
\no
\eea
The leading asymptotics depends on the range chosen for the value of the characteristics.
To obtain the asymptotics in as simple a manner as possible, we choose the integration
ranges for the torus to be $-\half \leq x_I', x_I'' \leq \half$. With this choice, it is the term 
$n=0$ which dominates,   and we find the following asymptotics, to leading order,
\bea
\label{6b2}
\tet [x](0,\Omega)= C_0 \, \tet ( \tau_1 x_1 + \tau x_2 + y_1 , \tau_1)
\eea
The term in $\tau x_2$ may be decomposed by making the following change of variables,
\bea
\label{6b3}
x_1 & \to & \tilde x_1 = x_1 + x_2 ( \Im \tau) / ( \Im \tau_1)
\no \\
y_1 & \to & \tilde y_1 = y_1 + x_2 \Big  ( \Re (\tau) - ( \Im \tau)  ( \Re \tau_1) / (\Im \tau_1)  \Big )
\eea
in terms of which the leading asymptotics of $\ln |\tet|^2$ becomes, 
\bea
\label{6b4}
\ln | \tet [x](0,\Omega)|^2 = \ln | \tet [\tilde x_1 ~ \tilde y_1] (0,\tau_1) |^2 - 2 \pi x_2 ^2 \, { \det Y \over \Im \tau_1}
\eea
where $\det Y = \det \Im \Omega = (\Im \tau_1 ) (\Im \tau_2) - ( \Im \tau)^2$.
Using translation invariance of the integration measure over the torus $T^4$, we have $dx_1 dy_1 = 
d\tilde x_1 d \tilde y_1$ and the integration range is unchanged by periodicity. Carrying out the 
integrations over $x_2$ and $y_2$, we find, 
\bea
\label{6b5}
\ln \Phi = \int _{-\half }^\half  d\tilde x_1 \int _{-\half }  ^\half d \tilde y_1 
\left (  \ln | \tet [\tilde x_1 ~ \tilde y_1] (0,\tau_1) |^2 - { \pi \over 6} \, { \det Y \over \Im \tau_1} \right )
\eea
Using the result of (\ref{6a5}) and (\ref{6a6}) for the remaining integral, we find,
\bea
\label{6b6}
\ln \Phi = \ln  | \eta (\tau_1)|^2 - { \pi \over 6} \, \left ( \Im \tau_2 - { (\Im \tau)^2 \over \Im \tau_1} \right )
\eea
Finally, we shall  work out the orders of the terms that we have omitted by retaining
only  the leading order in $\tau_2 \to i \infty$. We use the decomposition of  the 
genus-two $\tet$-function in (\ref{6b1}), still for the characteristics $-\half \leq x_I, y_I \leq \half$. 
The suppression factors are governed by the ratio of the correction terms, 
divided by the leading term, and take the form, 
\bea
\left | {C_n \over C_0} \right | = \exp 2 \pi \Big \{   - (\Im \tau _2) (n^2 + 2n x_2) - 2 \Im (\tau) x_1 n  \Big \}
\eea
Since $n^2 + 2 n x_2 \geq |n| ( |n| - 2 |x_2|)$, the contributions with $|n| \geq 2$ 
are suppressed by at least positive integer powers of $e^{ - 2 \pi \Im (\tau_2) }$. For $n=\pm 1$, 
the suppression is lesser, and is given by 
\bea
\exp 2 \pi \Big \{ - (\Im \tau_2) (1\pm 2 x_2) \mp 2 (\Im \tau) x_1 \Big \} 
{ \tet (x_1 \tau_1 + (x_2\pm 1) \tau +y_1, \tau_1) \over \tet (x_1 \tau_1 + x_2 \tau +y_1, \tau_1)}
\eea
Upon integration over $x_2$ in the range $-\half \leq x_2 \leq \half$, the correction is found to
be power law suppressed by  a factor $(\Im \tau_2)^{-1}$. Thus, the terms we have neglected 
are either exponentially suppressed for $|n| \geq 2$ and power suppressed for $|n|=1$. 

\subsection{The genus-one Arakelov Green function}

The genus-one Green function is standard up to normalization choices.  
The canonically normalized Green function $G$ was given in  (\ref{newprop}) while  the 
Arakelov normalization is obtained
by fixing the arbitrary additive constant in the Green function so that the integral
of the Arakelov Green function $\ln g$ vanishes. In the notation of (\ref{3b1}), we find 
\bea
\label{arakelovnorm}
\ln g (z) = \ln \left | { \tet _1 (z,\tau) \over \eta (\tau)} \right | - { \pi  ( \Im z  )^2 \over \Im \tau} 
\eea
We recall the product formula for $\tet_1 (z, \tau)$, 
\bea
\label{thetadef}
\vartheta_1(z,\tau) = -2 e^{i \pi \tau/4} \sin(\pi z)\prod_{m=1}^\infty 
\left( 1 - e^{2\pi i m \tau} \right) \left( 1 - e^{2\pi i m \tau+ 2 \pi i z} \right) \left( 1 - e^{2\pi i m \tau - 2 \pi i z} \right) 
\eea
The Arakelov Green function satisfies,
\bea
\label{intcon}
\int _\Sigma d^2 z \ln g(z-w)=0
\eea
The integral is over the fundamental domain of the genus-one surface $\Sigma$ 
generated by the lattice with periods $1$ and $\tau_1$. Note that  the canonical 
Green function $G$ in (\ref{newprop}) and the Arakelov Green function  $\ln g$ 
in (\ref{arakelovnorm})  are related by 
\bea
\label{genusonedef}
G(z,w) = -2 \ln g(z-w) + 2\ln(2\pi |\eta|^2)\,,
\eea
where we have used the fact that $\tet '(0,\tau)= -2\pi \eta(\tau)^3$.

\section{Direct calculation of separating degeneration of $\varphi$}
\label{secC}
\setcounter{equation}{0}

We calculate the separating degeneration asymptotics of $\varphi$ directly from the 
 formula for $\varphi$ given in (\ref{3d4}), with $G$ defined in (\ref{2a9}), and  $P$ 
defined in  (\ref{3a4}). In the parametrization of (\ref{6a0}), we seek the 
asymptotics as $\tau \to 0$ while keeping $\tau_I, ~ I=1,2$ fixed. The surface
$\Sigma$ pinches off to the union of two genus 1 surfaces $\Sigma _I$ 
with one puncture $p_I$ each. 

\sm

The leading asymptotics of the normalized holomorphic 
Abelian differentials $\omega ^t_I$ on the genus-two surface is given by
$\omega ^t _I (z) = \omega _I^{(1)} (z) + \cO(t)$ for $z \in \Sigma _I$ and 
$\omega ^t _I (z) =  \cO(t)$ for $z \not \in \Sigma _I$, with $\tau= \pi i t /2 + \cO(t^2)$.
Here, $\omega _I^{(1)}$ are the normalized holomorphic Abelian differentials on the 
genus-one components $\Sigma _I$. Representing $\Sigma _I$ by a flat torus with 
modulus $\tau_I$, and complex coordinates $z_I, \bar z_I$ with periods $1$ and $ \tau_I$, 
we have   $\omega_I ^{(1)}(z)=dz_I$. The imaginary part of the period matrix becomes, 
$Y_{IJ} = \delta _{IJ} \Im \tau_I + \cO(t)$. Using this information, we evaluate
the degeneration limit of the form $P(x,y)$ of (\ref{3a4}) and we find to leading order,
\bea
\label{C1}
z, w \in \Sigma _I ~~~ & \hskip 0.4in & P(z,w) = - 4 (\Im \tau_I)^{-2}  d^2z d^2w
\no \\
z\in \Sigma _1, ~ w \in \Sigma _2 && P(z,w) = 4 (\Im \tau_1)^{-1} (\Im \tau_2)^{-1} d^2z d^2w
\eea
where we recall that $d^2 z = i dz \wedge d \bar z/2$, so that 
$\int _{\Sigma _I} d^2 z_I =\Im \tau_I$.
The asymptotics of the Green function $G$ was 
carefully evaluated in  formula (3.19) of \cite{D'Hoker:2005ht}, and is given by,
\bea
\label{C2}
z, w \in \Sigma _I ~~~ & \hskip 0.4in & G(z,w) = G^{(1)} (z,w; \tau_I) + \cO( \tau)
\\
z\in \Sigma _1, ~ w \in \Sigma _2 && G(z,w) = 
\ln { |\tau | \over 2 \pi} + G^{(1)} (z, p_1, \tau_1) + G^{(1)} (w, p_2; \tau_2) + \cO (\tau)
\no
\eea
To make its genus-one nature and modulus explicit,  we have denoted the genus-one 
Green function of (\ref{2a10}) for modulus $\tau_I$ by $G^{(1)} (z,w; \tau_I)$. 

\sm

To carry out the integrals of (\ref{3d4}) in this limit, we proceed from (\ref{intcon}) and (\ref{genusonedef}), to deduce the value of the following integral,
\bea
\label{C3}
\int _{\Sigma _I} d^2 z_I G^{(1)} (z_I, w; \tau_I) = 2 (\Im \tau_I) \ln ( 2 \pi |\eta (\tau_I)|^2)
\eea
for any point $w$ (as follows by translation invariance on the torus).
Combining the limits of $P$ and $G$ given above, and carrying out the integrals over
$z$ and $w$, we find,
\bea
\varphi & = &  + \ln ( 2 \pi |\eta (\tau_1)|^2)  + \ln ( 2 \pi |\eta (\tau_2)|^2) 
\no \\ &&
- \ln { |\tau | \over 2 \pi} -2  \ln ( 2 \pi |\eta (\tau_1)|^2)  -2 \ln ( 2 \pi |\eta (\tau_2)|^2) 
\eea
The terms on the first line result from the integration over $z,w$ on the same
component $\Sigma _I$, while the terms on the second line arise from $z$ and $w$ on opposite 
components $\Sigma _I$. The contribution from points in the funnel (present for non-zero $t$)
tends to 0 as $t \to 0$, and may be neglected. Minor simplification  reproduces
the separating degeneration asymptotics of (\ref{6a8}), which was obtained through
the asymptotics from $\Phi$ in appendix B.


\newpage

\begin{thebibliography}{123}
\itemsep=0in

\bibitem{D'Hoker:2001nj} 
  E.~D'Hoker and D.~H.~Phong,
  ``Two loop superstrings. 1. Main formulas,''
  Phys.\ Lett.\ B {\bf 529}, 241 (2002)
  [hep-th/0110247].

\bibitem{D'Hoker:2005jc} 
  E.~D'Hoker and D.~H.~Phong,
  ``Two-loop superstrings VI: Non-renormalization theorems and the 4-point function,''
  Nucl.\ Phys.\ B {\bf 715}, 3 (2005)
  [hep-th/0501197].

\bibitem{D'Hoker:2002gw} 
  E.~D'Hoker and D.~H.~Phong,
  ``Lectures on two loop superstrings,''
  Conf.\ Proc.\ C {\bf 0208124}, 85 (2002)
  [hep-th/0211111].


\bibitem{Berkovits:2005ng} 
  N.~Berkovits and C.~R.~Mafra,
  ``Equivalence of two-loop superstring amplitudes in the pure spinor and RNS formalisms,''
  Phys.\ Rev.\ Lett.\  {\bf 96}, 011602 (2006)
  [hep-th/0509234].

\bibitem{Green:1997tv}
  M.~B.~Green and M.~Gutperle,
  ``Effects of D instantons,''
  Nucl.\ Phys.\ B {\bf 498} (1997) 195
  [hep-th/9701093].
    
\bibitem{Green:1998by}
  M.~B.~Green and S.~Sethi,
  ``Supersymmetry constraints on type IIB supergravity,''
  Phys.\ Rev.\ D {\bf 59} (1999) 046006
  [hep-th/9808061].

  
\bibitem{Sinha:2002zr}
  A.~Sinha,
  ``The $\hat G^4 \lambda^{16}$ term in IIB supergravity,''
  JHEP {\bf 0208} (2002) 017
  [hep-th/0207070].

\bibitem{Green:2005ba}
M.B. Green and  P. Vanhove, 
{\sl Duality and higher
  derivative terms in M theory},
JHEP {\bf 0601} (2006) 093
  [arXiv:hep-th/0510027].

  
  
\bibitem{Zhang}
S.W. Zhang, ``Gross - Schoen Cycles and Dualising Sheaves"
Inventiones mathematicae, Volume 179, Issue 1, pp 1-73
arXiv:0812.0371

\bibitem{Kawazumi}
N. Kawazumi, ``Johnson's homomorphisms and the Arakelov Green function",
arXiv:0801.4218 [math.GT].


\bibitem{DeJong2}
R. De Jong,
``Second variation of Zhang's $\lambda$-invariant on the moduli space of curves",
arXiv:1002.1618v2.


\bibitem{Bost1}
J.-B. Bost, ``Fonctions de Green-Arakelov, fonctions theta et courbes de genre 2",
C. R. Acad. Sci. Paris, t. 305, s\'erie I, p. 643-646, 1987

\bibitem{Bost2}
J.-B. Bost, J.-F. Mestre, L. Morey-Bailly, ``Sur le calcul explicite des classes de Chern des
surfaces arithm\'etiques de genre 2", in {\sl S\'eminaire sur les pinceaux de courbes elliptiques},
Ast\'erique 183 (1990) 69-105.

\bibitem{D'Hoker:2005ht}
  E.~D'Hoker, M.~Gutperle and D.~H.~Phong,
  ``Two-loop superstrings and S-duality,''
  Nucl.\ Phys.\  B {\bf 722} (2005) 81
  [arXiv:hep-th/0503180].

\bibitem{DeJong1}
R. De Jong, ``Asymptotic behavior of the Kawazumi-Zhang invariant for degenerating Riemann surfaces",
arXiv:1207.2353 



\bibitem{Green:1987sp}
  M.~B.~Green, J.~H.~Schwarz and E.~Witten,
  {\sl Superstring Theory. Vol.~1: Introduction,}
{\it  Cambridge, Uk: Univ. Pr. (1987)  (Cambridge Monographs On Mathematical Physics)}

\bibitem{Klingen}
H. Klingen, {\sl Introductory Lectures to Siegel Modular Forms},
Cambridge University Press, 1990

\bibitem{Green:1999pv}
  M.~B.~Green and P.~Vanhove,
  ``The Low-energy expansion of the one loop type II superstring amplitude,''
  Phys.\ Rev.\ D {\bf 61} (2000) 104011
  [hep-th/9910056].
  
  
\bibitem{Green:2008uj}
  M.~B.~Green, J.~G.~Russo and P.~Vanhove,
  ``Low energy expansion of the four-particle genus-one amplitude in type II superstring theory,''
  JHEP {\bf 0802} (2008) 020
  [arXiv:0801.0322 [hep-th]].


\bibitem{Green:1981yb} 
  M.~B.~Green and J.~H.~Schwarz,
``Supersymmetrical String Theories,''
  Phys.\ Lett.\ B {\bf 109}, 444 (1982).

\bibitem{D'Hoker:1994yr} 
  E.~D'Hoker and D.~H.~Phong,
``The Box graph in superstring theory,''
  Nucl.\ Phys.\ B {\bf 440}, 24 (1995)
  [hep-th/9410152].


\bibitem{Schlotterer:2012ny} 
  O.~Schlotterer and S.~Stieberger,
  ``Motivic Multiple Zeta Values and Superstring Amplitudes,''
  arXiv:1205.1516 [hep-th].


\bibitem{Green:2013bza}
  M.~B.~Green, C.~R.~Mafra and O.~Schlotterer,
  ``Multiparticle one-loop amplitudes and S-duality in closed superstring theory,''
  arXiv:1307.3534 [hep-th].


\bibitem{DeJong3}
R. De Jong, ``Arakelov Invariants of  Riemann surfaces",
Documenta Mathematica 10 (2005) 311Ð329


\bibitem{Wentworth:1991}
R. Wentworth  
``The Asymptotics of the Arakelov-Green's Function and Faltings' Delta Invariant'',
Commun. Math. Phys. 137, 427-459 (1991)

\bibitem{DM}
P. Deligne and D. Mumford, Publ. Math. IHES {\bf 36} (1979) page 75.

\bibitem{Obers:1999um} 
  N.~A.~Obers and B.~Pioline,
  ``Eisenstein series and string thresholds,''
  Commun.\ Math.\ Phys.\  {\bf 209}, 275 (2000)
  [hep-th/9903113].

  
\bibitem{Green:2010wi}
  M.~B.~Green, J.~G.~Russo and P.~Vanhove,
  ``Automorphic properties of low energy string amplitudes in various dimensions,''
  Phys.\ Rev.\ D {\bf 81} (2010) 086008
  [arXiv:1001.2535 [hep-th]].
  

\bibitem{Green:2010kv}
  M.~B.~Green, S.~D.~Miller, J.~G.~Russo and P.~Vanhove,
  ``Eisenstein series for higher-rank groups and string theory amplitudes,''
  Commun.\ Num.\ Theor.\ Phys.\  {\bf 4} (2010) 551
  [arXiv:1004.0163 [hep-th]].

\bibitem{Pioline:2010kb} 
  B.~Pioline,
  ``R**4 couplings and automorphic unipotent representations,''
  JHEP {\bf 1003}, 116 (2010)
  [arXiv:1001.3647 [hep-th]].
  


\bibitem{D'Hoker:2001qp} 
  E.~D'Hoker and D.~H.~Phong,
  ``Two loop superstrings 4: The Cosmological constant and modular forms,''
  Nucl.\ Phys.\ B {\bf 639}, 129 (2002)
  [hep-th/0111040].
  



\end{thebibliography}
\end{document}